\documentclass[twocolumn,nofootinbib,superscriptaddress,preprintnumbers,floatfix]{revtex4}

\usepackage[small]{subfigure}
\usepackage[bookmarks=false,hyperfootnotes=false]{hyperref}
\usepackage{graphicx}
\usepackage{amsmath}

\newcommand{\eq}[1]{Eq.~\eqref{eq:#1}}
\newcommand{\eqs}[2]{Eqs.~\eqref{eq:#1} and \eqref{eq:#2}}
\renewcommand{\sec}[1]{Sec.~\ref{sec:#1}}

\newcommand{\subsec}[1]{Sec.~\ref{subsec:#1}}
\newcommand{\fig}[1]{Fig.~\ref{fig:#1}}
\newcommand{\figs}[2]{Figs.~\ref{fig:#1} and \ref{fig:#2}}
\newcommand{\app}[1]{App.~\ref{app:#1}}

\newcommand{\abs}[1]{\lvert#1\rvert}
\newcommand{\ord}[1]{\mathcal{O}(#1)}
\newcommand{\MAe}[3]{\Bigl\langle#1\Bigl\lvert#2\Bigr\rvert#3\Bigr\rangle}

\newcommand{\df}{\mathrm{d}}
\newcommand{\img}{\mathrm{i}}

\renewcommand{\Re}{\mathrm{Re}}
\newcommand{\Li}{\mathrm{Li}}

\newcommand{\tr}{\mathrm{tr}}

\newcommand{\eps}{\epsilon}

\newcommand{\Tau}{\mathcal{T}}

\newcommand{\cJ}{{\mathcal J}}
\newcommand{\cL}{{\mathcal L}}

\newcommand{\bT}{\mathbf{T}}

\newcommand{\vC}{\vec{C}}
\newcommand{\vT}{\vec{T}}

\newcommand{\hq}{\hat{q}}
\newcommand{\hs}{\hat{s}}
\newcommand{\hH}{\widehat{H}}
\newcommand{\hS}{\widehat{S}}
\newcommand{\hT}{\widehat{T}}
\newcommand{\hZ}{\widehat{Z}}
\newcommand{\hga}{\widehat{\gamma}}

\newcommand{\nn}{\nonumber}

\newcommand{\Ecm}{E_\mathrm{cm}}
\newcommand{\cusp}{\mathrm{cusp}}
\newcommand{\bare}{\mathrm{bare}}
\newcommand{\hemi}{\mathrm{hemi}}
\newcommand{\cut}{\mathrm{cut}}

\newcommand{\one}{{(1)}}

\newcommand{\id}{\mathbf{1}}

\allowdisplaybreaks[4]

\begin{document}


\preprint{MIT--CTP 4214}

\title{\boldmath The Soft Function for Exclusive $N$-Jet Production at Hadron Colliders}

\author{Teppo T.~Jouttenus}
\affiliation{Center for Theoretical Physics, Massachusetts Institute of Technology, Cambridge, MA 02139, USA\vspace{0.5ex}}

\author{Iain W.~Stewart}
\affiliation{Center for Theoretical Physics, Massachusetts Institute of Technology, Cambridge, MA 02139, USA\vspace{0.5ex}}
\affiliation{Center for the Fundamental Laws of Nature, Harvard University, Cambridge, MA 02138, USA\vspace{0.5ex}}

\author{Frank J.~Tackmann}
\affiliation{Center for Theoretical Physics, Massachusetts Institute of Technology, Cambridge, MA 02139, USA\vspace{0.5ex}}

\author{Wouter J.~Waalewijn\vspace{0.5ex}}
\affiliation{Department of Physics, University of California at San Diego, La Jolla, CA 92093, USA\vspace{0.5ex}}

\date{February 21, 2011}

\begin{abstract}

  The $N$-jettiness event shape divides phase space into $N+2$ regions, each
  containing one jet or beam. Using a geometric measure these regions correspond
  to jets with circular boundaries.  We give a factorization theorem for the
  cross section fully differential in the mass of each jet, and
  compute the corresponding soft function at next-to-leading order (NLO). The
  ultraviolet divergences are analytically extracted by exploiting hemispheres
  for interactions between each pair of hard partons, leaving only convergent
  integrals that are sensitive to the precise boundaries. This hemisphere decomposition
  can also be applied to other $N$-jet soft functions, including other observables.
  For $N$-jettiness, the final result for the soft
  function involves stable one-dimensional numerical integrals, and all
  ingredients are now available to extend NLO cross sections to resummed
  predictions at next-to-next-to-leading logarithmic order.

\end{abstract}

\maketitle

\section{Introduction}
\label{sec:intro}

The measurement of exclusive jet cross sections, where one identifies a certain
number of signal jets but vetoes additional jets, is an important aspect of
Higgs and new-physics searches at the LHC and Tevatron. In such searches, the
experiments often analyze the data separated into bins of different numbers of
jets. This is done because the relative contributions of various signal and
background channels often vary with the number of hard jets in the event. Hence,
it is important to optimize the analyses for each jet bin. An important example
is the Higgs search at the Tevatron~\cite{:2010ar}, which analyzes the data
separately for $H + 0$ jets, $H + 1$ jet, and $H + 2$ or more jets.

Reliable theoretical calculations of exclusive jet cross
sections are of course essential. The complication compared to the calculation of
an inclusive $N$-jet cross section, where one sums over any additional jets,
comes from the fact that the veto on additional jets imposes a restriction on
the energetic initial- and final-state radiation off the primary hard
partons, as well as the overall soft radiation in the event. This restriction on
additional emissions leads to the appearance of large Sudakov double logarithms
in the perturbation theory. This is a well-known phenomenon and is due to an
incomplete cancellation of infrared contributions between virtual corrections
and restricted real radiation. For this reason, the calculation of exclusive jet
cross sections is traditionally carried out with parton-shower Monte Carlo
programs, where the parton shower allows one to resum the most singular leading
double logarithms.

An alternative analytic approach to calculate exclusive jet cross sections is
possible using factorization and the methods of soft-collinear effective theory
(SCET)~\cite{Bauer:2000ew,Bauer:2000yr,Bauer:2001ct,Bauer:2001yt}.  SCET is
designed to study processes with a specific number of hard jets. It allows one
to factorize the $N$-jet cross section into individually calculable pieces and
resum the large logarithmic contributions to obtain a convergent perturbative
series. The advantage of this approach is that the resummation can be carried
out to much higher orders than is possible with parton showers. In addition, it
is much easier than in a parton-shower program to include higher-order virtual
corrections, and to correctly reproduce the inclusive cross section in the limit
when the jet-veto cut is eliminated.

Schematically, the cross section for $pp \to N$ jets (plus some nonhadronic
final state which we suppress for now) can be factorized as~\cite{Bauer:2002nz,
Bauer:2008jx, Stewart:2009yx}
\begin{equation} \label{eq:sigmaN}
\sigma_N = H_N \times \Bigl[ B_a B_b \prod_{i = 1}^N J_i \Bigr] \otimes S_N
\,.\end{equation}
This formula directly applies to observables that implement a veto on additional
jets which restricts the phase space to the exclusive $N$-jet region (assuming
that Glauber effects cancel as they do in Drell-Yan~\cite{Collins:1988ig}). The
hard function $H_N$ encodes hard virtual corrections to the underlying partonic
$2 \to N$ process, the beam functions $B_{a,b}$ contain the parton distributions
and perturbative collinear initial-state radiation from the colliding hard
partons, and the jet functions $J_i$ describe energetic collinear final-state
radiation from the primary $N$ hard partons produced in the collision. The soft
function $S_N$ describes the soft radiation in the event that couples to the in-
and outgoing hard partons. Since the collinear and soft radiation are not
separately observable, the soft function is convolved with the beam and jet
functions. The veto on additional jets restricts the collinear initial-state radiation, the final-state radiation, and
the soft radiation, which means the precise definition of the required beam,
jet, and soft functions depends on the veto variable.

For the case of an exclusive $0$-jet cross section, inclusive beam and jet
functions can be obtained by using a simple event-shape variable called beam
thrust~\cite{Stewart:2009yx} to veto central jets. This $0$-jet cross section
has been studied for Drell-Yan production in Ref.~\cite{Stewart:2010pd} and for
Higgs production in Ref.~\cite{Berger:2010xi}. The latter is for example
relevant for the $H\to WW^*$ search channel, where a jet veto is needed to
remove the large $t \bar t \to WW b \bar b$ background. The use of an event
shape for the jet veto makes a resummation of large logarithms to next-to-next-to-leading logarithmic (NNLL) order possible.

The generalization of beam thrust to processes with $N$ jets is $N$-jettiness,
$\Tau_N$, introduced in Ref.~\cite{Stewart:2010tn}. It is designed such that in
the limit $\Tau_N \to 0$ the final state consists of $N$ narrow jets plus two
narrow initial-state radiation jets along the beam axis (for hadron collisions).  Since it does not
restrict the collinear radiation inside a jet, the beam and jet functions
appearing in~\eq{sigmaN} are again the inclusive beam and jet functions (which
are known to one~\cite{Fleming:2006cd, Stewart:2010pd, Mantry:2009qz,
  Berger:2010xi} and two loops~\cite{Becher:2006qw, Becher:2010pd},
respectively). Furthermore, since $N$-jettiness itself covers all of phase
space, no additional restriction on the radiation outside of jets or beams
is needed. In contrast, hadron-collider event shapes constructed from transverse
momenta only, such as transverse thrust, in general require the addition
of exponentially suppressed forward terms to suppress the contributions from
large rapidities~\cite{Banfi:2004nk, Banfi:2010xy}.

Factorization for $N$-jettiness can be contrasted with factorization for jet algorithms. Here, the
perturbative corrections are complicated by: the presence of non-global
logarithms~\cite{Dasgupta:2001sh,Banfi:2002hw,Dokshitzer:2003uw,Banfi:2010pa},
the potential for soft radiation to be strongly influenced by the number of
energetic partons in the jets, and by cuts on soft radiation that introduce
additional soft scales that must be handled within
factorization~\cite{Ellis:2009wj,Ellis:2010rwa}.  Jet functions for jet
algorithms in $e^+e^- \to $ jets have been calculated at next-to-leading order
(NLO) in Refs.~\cite{Jouttenus:2009ns, Ellis:2010rwa}.  In
Ref.~\cite{Ellis:2010rwa} the soft function for $e^+e^-\to $ jets was calculated
at NLO, where a cut on the total energy outside the jets was used as the jet
veto. Using $N$-jettiness avoids these issues that complicate the
structure of perturbation theory.

The $N$-jettiness event shape assigns all particles to one of $N+2$ regions,
corresponding to the $N$ jets and $2$ beams. Therefore $\Tau_N$ acts much like a
jet algorithm, and we can consider distinct measurements on each of these
``jets''. The simplest example is $\Tau_N^i$, the $N$-jettiness contribution
from each region $i$, where $\Tau_N = \sum_i \Tau_N^i$.  A measurement of
$\Tau_N^i$ is essentially the same as measuring the transverse mass or invariant
mass of this jet.
This correspondence will be made precise in the next section. We will also
briefly explore the shape of the jet regions obtained using $N$-jettiness with
different measures. A geometric measure gives jets with circular boundaries,
putting them in the class that are typically preferred experimentally.

For an $N$-jettiness cross section calculation using \eq{sigmaN}, the only
missing ingredient for an evaluation of generic processes at NNLL is the
one-loop $N$-jettiness soft function, $S_N$, which we compute in detail in this
paper.  (As mentioned above, the beam and jet functions are known. The hard
function in \eq{sigmaN} can be obtained from the corresponding QCD fixed-order
calculation, many of which are now known to NLO.)  General features of
$N$-jettiness and its jet regions are explored in \sec{general}. Results are
given for the fully differential $\Tau_N^i$ factorization theorem, and for
renormalization group consistency equations for the $N$-jettiness soft function.
Section~\ref{sec:1jet} contains details of the NLO calculation of $S_N$,
including developing a simple method that uses hemispheres for each pair of hard
partons to extract UV divergences and the corresponding induced logarithmic
terms.  The remaining $\ord{\alpha_s}$ terms are then given by finite integrals
that do not involve the UV
regulator, and we will refer to these as the non-hemisphere contributions.
This hemisphere decomposition is not specific to the $N$-jettiness observable, and we show
how it can be applied in general. For the $N$-jettiness soft function we reduce the
non-hemisphere contributions to well-behaved one-dimensional numerical integrals
(some details are relegated to appendices).
Section~\ref{sec:conclusions} contains conclusions.

Although it is not directly related to our investigations here, it is worth
mentioning that $N$-jettiness is useful for exploring jet
substructure~\cite{Kim:2010uj,Thaler:2010tr}. This is done with
$N$-subjettiness, which restricts the definition of the event shape to particles
and reference momenta inside a jet. There are interesting correspondences
between applications of $N$-jettiness and $N$-subjettiness. In particular one
could study the mass spectrum of subjets with $\Tau_N^i$, following
a similar procedure that we advocate here for jets.

\section{Setup of the Calculation}
\label{sec:general}

\subsection{\boldmath $N$-Jettiness Definition and Regions}
\label{subsec:def}

\begin{figure*}[t!]
\includegraphics[width=0.33\textwidth]{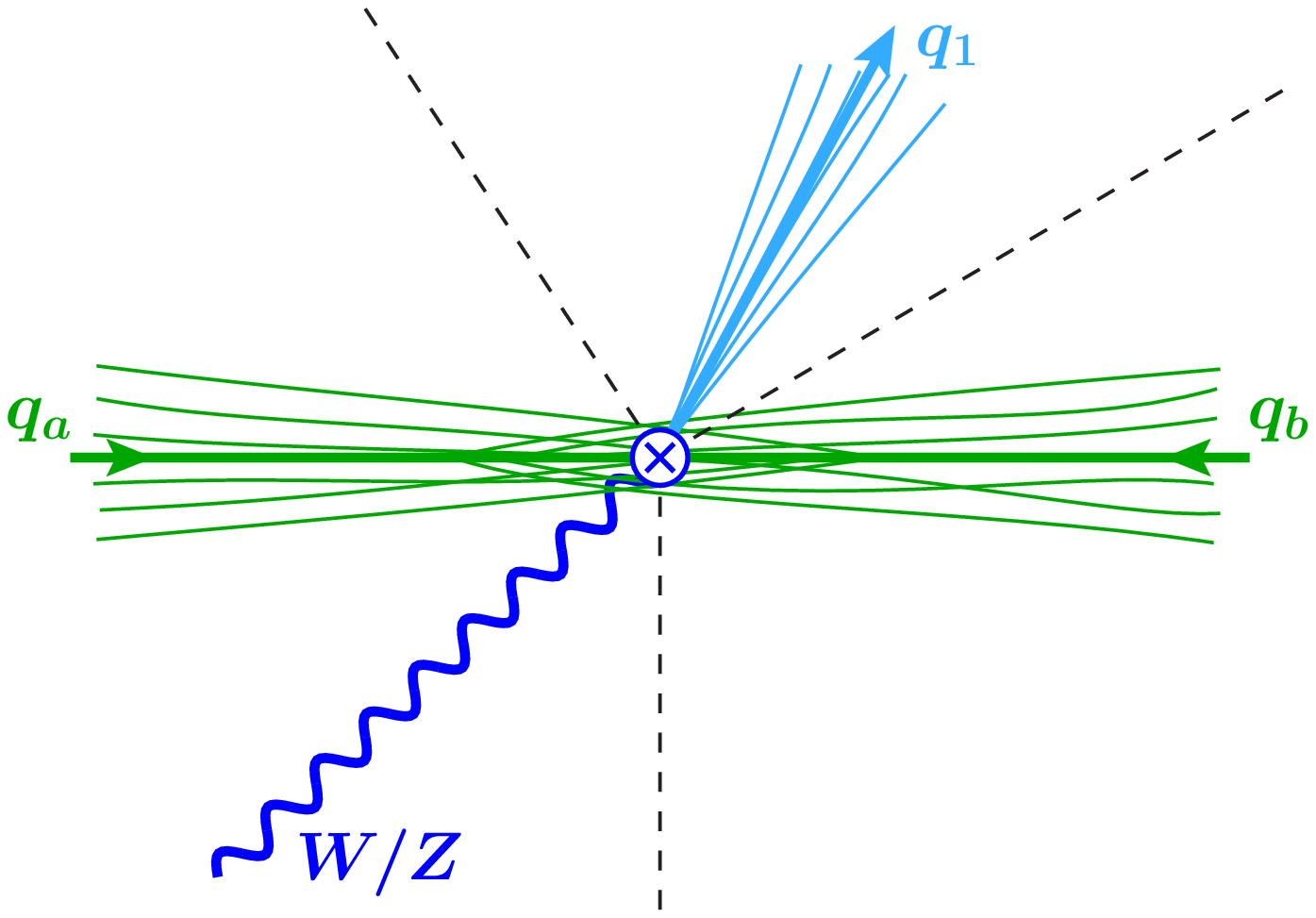}%
\hfill%
\includegraphics[width=0.33\textwidth]{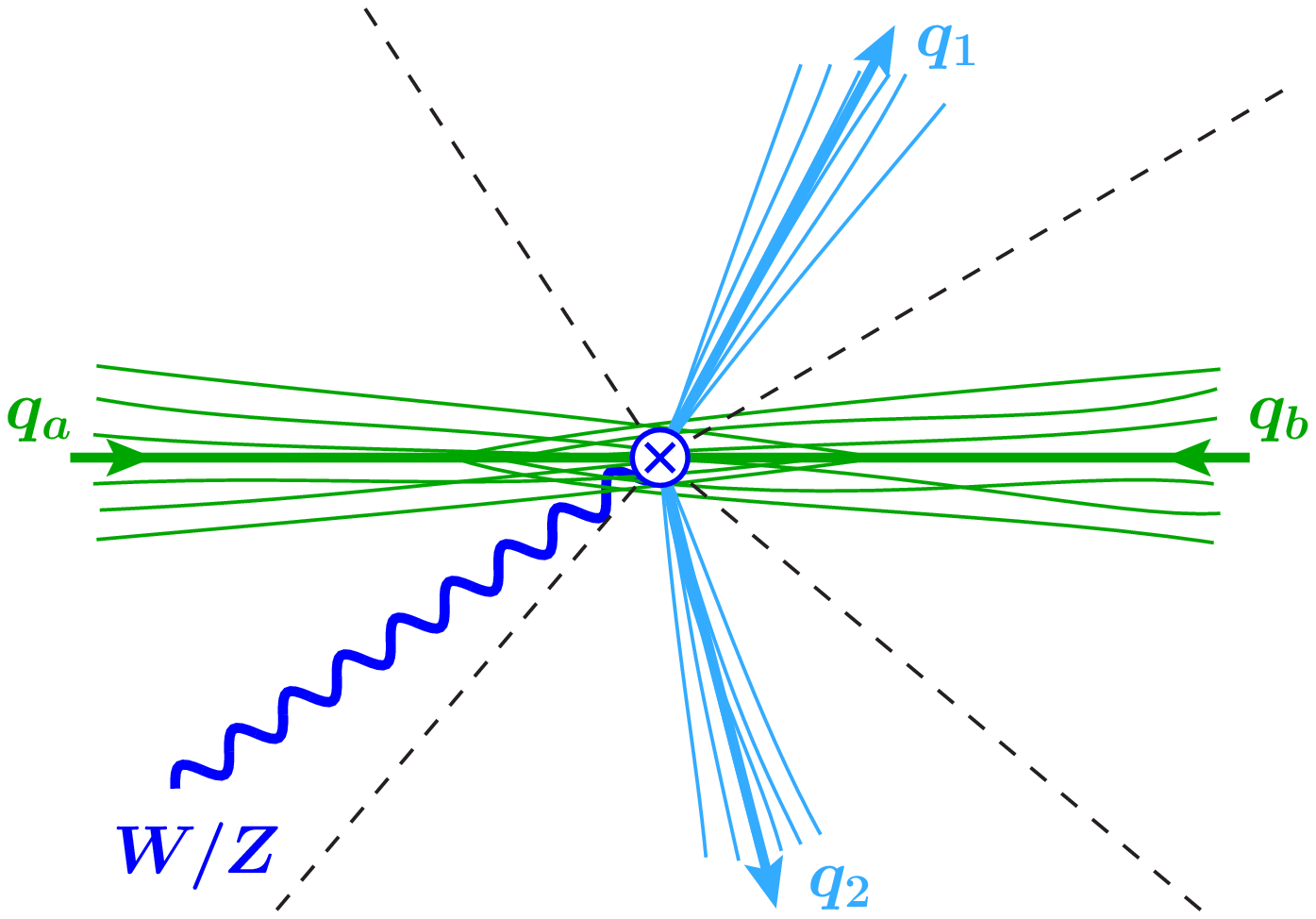}%
\hfill%
\includegraphics[width=0.33\textwidth]{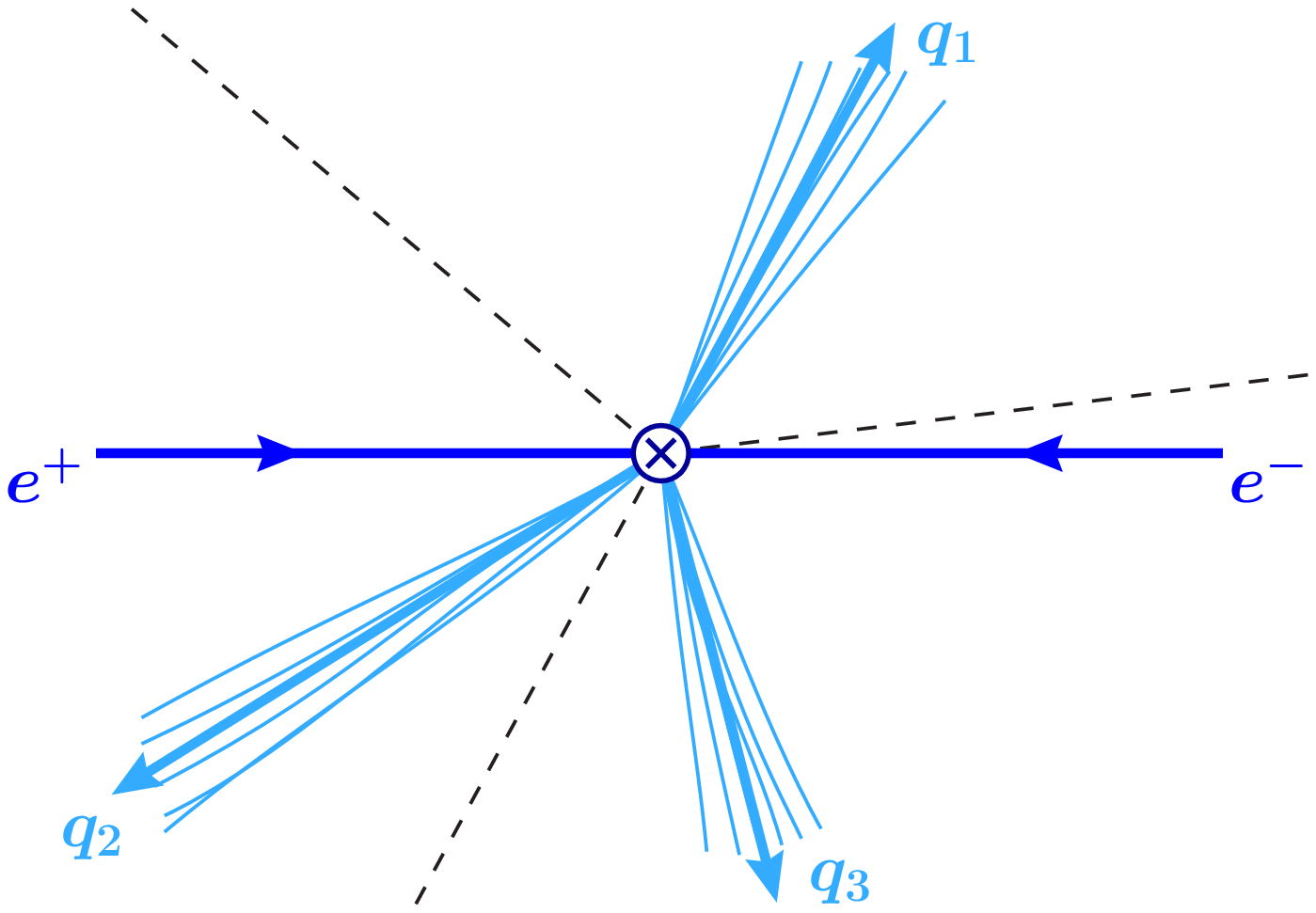}%
\vspace{-0.5ex}
\caption{ Jet and beam reference momenta for $1$-jettiness (left), $2$-jettiness (middle) and $e^+e^-$ $3$-jettiness (right). In the middle plot the jets and beams do not necessarily lie in a plane.}
\label{fig:jettiness}
\end{figure*}

\begin{figure*}[t!]
\hfill%
\includegraphics[width=0.4\textwidth]{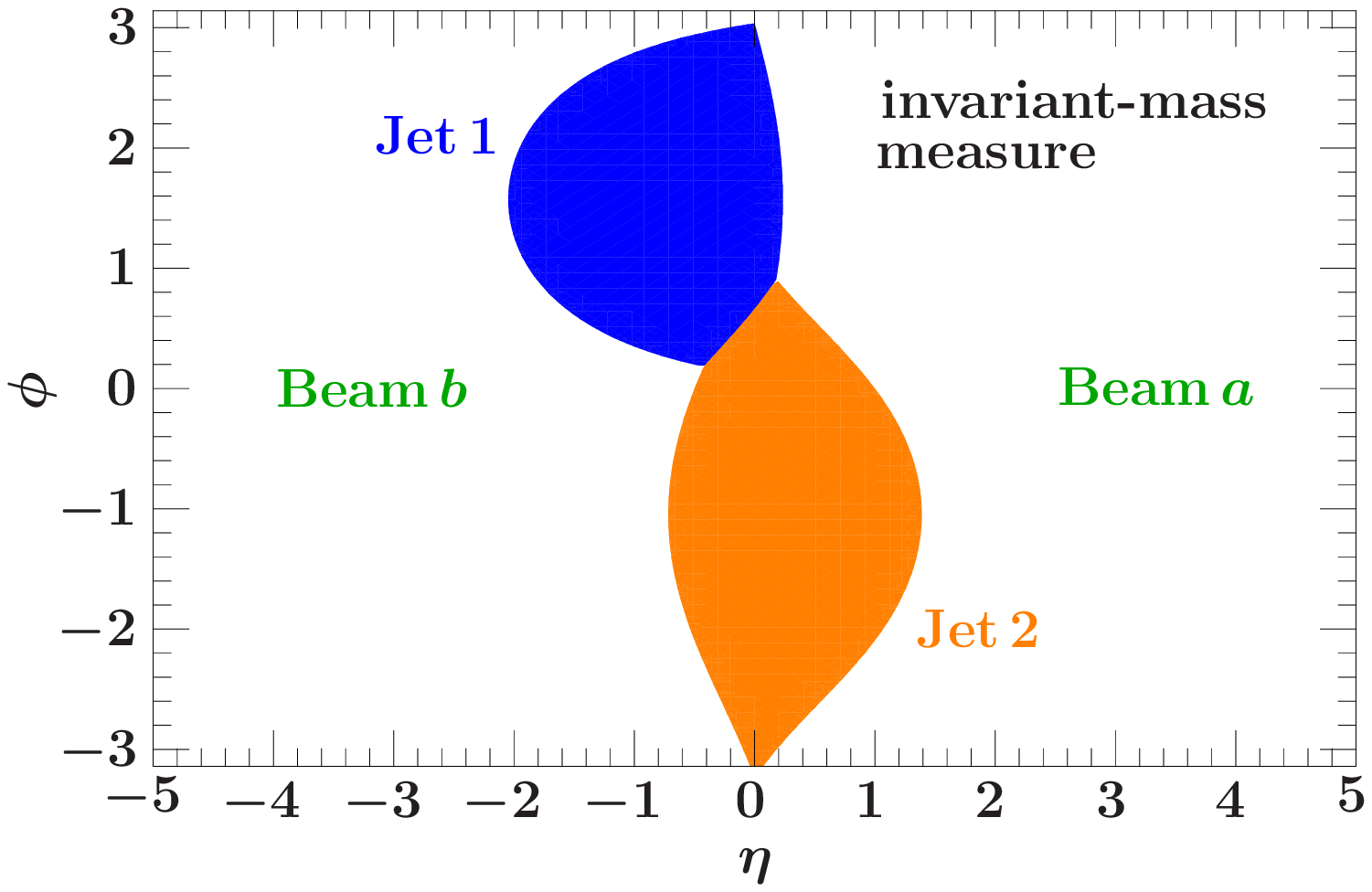}%
\hfill%
\includegraphics[width=0.4\textwidth]{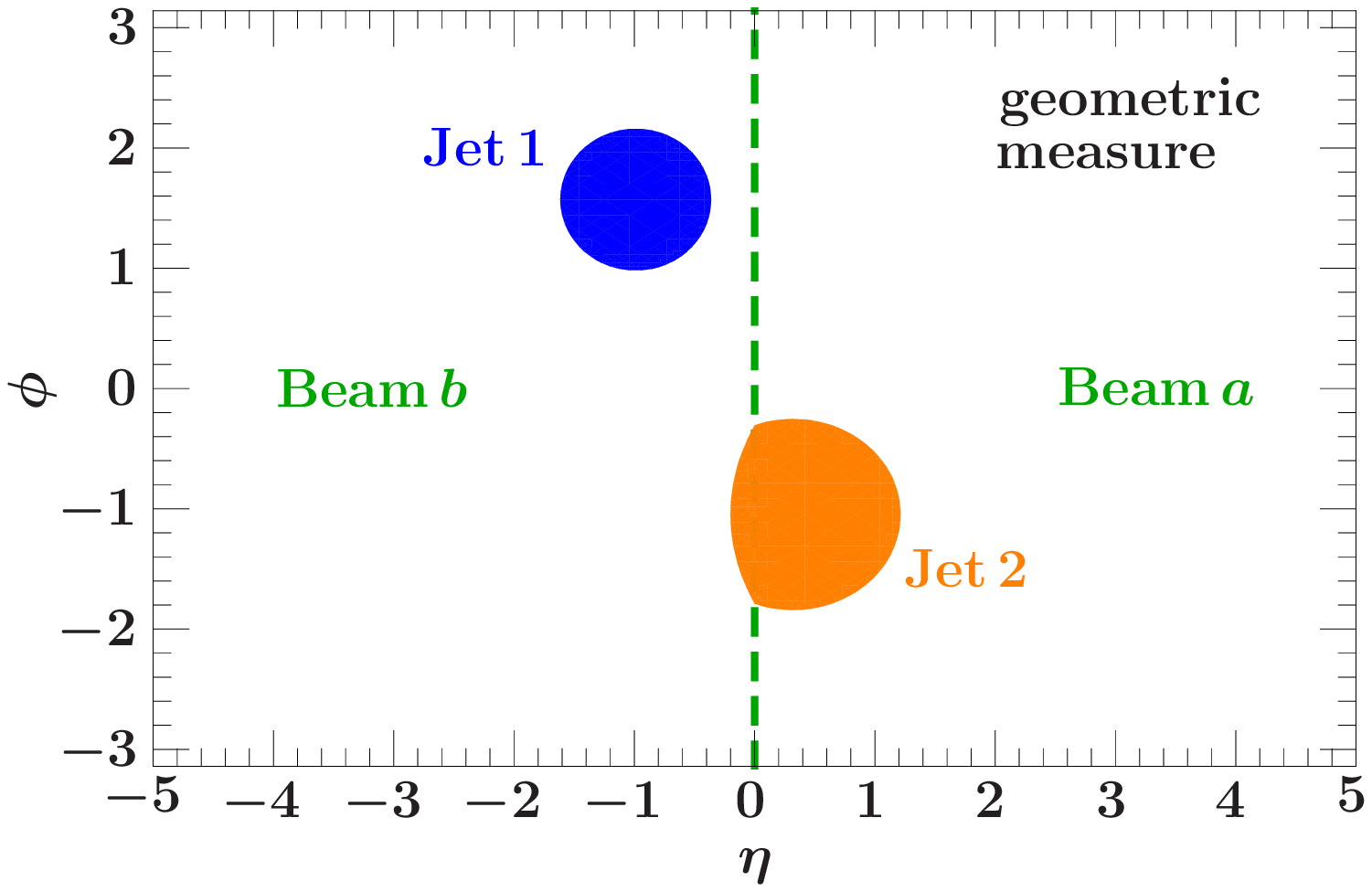}%
\hspace*{\fill}%
\vspace{-0.5ex}
\caption{The jet and beam regions for the same two jets using $2$-jettiness. On the left we use the invariant-mass measure $Q_i = Q$. On the right we use the geometric measure with $Q_i = \abs{\vec{q}_{iT}}$ for the jets and $Q_{a,b} = x_{a,b}\Ecm$ for the beams.}
\label{fig:2jet_etaphi}
\end{figure*}

$N$-jettiness is defined as~\cite{Stewart:2010tn}
\begin{equation}\label{eq:TauN_def}
\Tau_N = \sum_k
\min_i \Bigl\{ \frac{2 q_i\cdot p_k}{Q_i} \Bigr\}
\,,\end{equation}
where $i$ runs over $a, b$ for the two beams and $1, \ldots, N$ for the
final-state jets.%
\footnote{Here we use a dimension-one $\Tau_N$ as in Ref.~\cite{Berger:2010xi}
as opposed to the dimensionless $\tau_N$ of Ref.~\cite{Stewart:2010tn}.}
For $e^+e^-$ collisions, the terms for the beams are absent
and we continue to let $N$ refer to the number of jets.  The complexity of the
calculation for the $e^+e^-$ $(N+2)$-jettiness is equivalent to $N$-jettiness
for $pp$ collisions. In \eq{TauN_def} the $q_i$ are massless reference momenta
for the jets and beams, and the $Q_i$ are normalization factors. For each jet we
can take
\begin{equation}
q_i^\mu = \omega_i\, (1,\vec n_i)
\,,\end{equation}
where $\omega_i$ is the jet energy, and $\vec n_i$ is the jet direction. The
$\omega_i$ and $\vec n_i$ can be predetermined with a suitable jet algorithm,
and the choice of algorithm only gives power-suppressed effects, as explained in
Ref~\cite{Stewart:2010tn}.  For the beams we have
\begin{equation}
q_a^\mu = \frac12 x_a\, \Ecm (1, \hat z)
\,,\qquad
q_b^\mu = \frac12 x_b\, \Ecm (1, - \hat z)
\,,\end{equation}
where $\Ecm$ is the center-of-mass energy, $\hat z$ points along the beam axis,
and $x_{a,b}$ are the light-cone momentum fractions of the colliding hard
partons. The latter are defined as
\begin{equation}
x_a\Ecm = Q\,e^Y
\,,\qquad
x_b\Ecm = Q\,e^{-Y}
\,,\end{equation}
where $Q^2$ and $Y$ are the total invariant mass-squared and rapidity of the
hard interaction. They are determined from the observed final state by
\begin{align}
Q^2 &= x_a x_b \Ecm^2 = (q_1 + \dotsb + q_N + q)^2
\,,\nn\\
2 Y &= \ln\frac{x_a}{x_b} = \ln\frac{(1, -\hat z) \cdot (q_1 + \dotsb + q_N + q)}{(1, \hat z) \cdot (q_1 + \dotsb + q_N + q)}
\,.\end{align}
Here $q^\mu$ denotes the total momentum of the non-hadronic final state if one is present.

The choice of the $q_i^\mu$ is illustrated in \fig{jettiness} for $1$-jettiness (left panel), $2$-jettiness (middle panel), and $e^+e^-$ $3$-jettiness (right panel). For the first two cases $q^\mu$ is given by the momentum of the $W/Z$. In SCET the $q_i^\mu$'s become the large label momenta on the collinear fields, which can be thought of as the momenta of the partons in the hard interaction.
The minimum in \eq{TauN_def} divides the total phase space into $N+2$ regions, one for each beam and jet, as indicated by the dashed lines in \fig{jettiness}. Their union exactly covers all of phase space, and
the boundary between any two regions is a (part of a) cone.

The $Q_i$ in \eq{TauN_def} are dimension-one variables that characterize the
hardness of the jets. Different choices for the $Q_i$ correspond to
choosing different distance measures in the minimization in $\Tau_N$.  For
example, for fixed $Q_i = Q$, the distance measure is just the invariant mass,
$2q_i\cdot p_k$. The resulting jet and beam regions in this case are illustrated
for $2$-jettiness in the left panel of \fig{2jet_etaphi}.
Choosing the jet transverse momentum $Q_i = \abs{\vec{q}_{iT}}$
for the jets, the measure becomes a geometric measure, which is boost-invariant
along the beam axis,
\begin{align} \label{eq:geometric}
\frac{2 q_i\cdot p_k}{\abs{\vec{q}_{iT}}}
&= \abs{\vec p_{kT}}\, (2\cosh \Delta \eta_{ik} - 2\cos \Delta \phi_{ik})
\nn\\
&\approx \abs{\vec p_{kT}}\,\bigl[ (\Delta \eta_{ik})^2 + (\Delta \phi_{ik})^2 \bigr]
\,.\end{align}
Here, $\Delta\eta_{ik} = \eta_i - \eta_k$, $\Delta\phi_{ik} = \phi_i - \phi_k$
are the differences in (pseudo)rapidity and azimuthal angle between the
direction of jet $i$ and particle $k$.  The second line is valid in the limit of
small $\Delta\eta$ and $\Delta\phi$. Equation~\eqref{eq:geometric} results in circular
boundaries for the jet regions, as illustrated in the right panel of \fig{2jet_etaphi}.
In this case only the $\vec n_i$ part of
$q_i^\mu$ enters, and the $\vec n_i$ could be obtained by the choice which
minimizes $\Tau_N$, thus making $N$-jettiness a true event shape that does
not depend on any auxiliary input from a jet algorithm. The jet energy is then
simply given by summing over the particles in each jet region as determined by
$\Tau_N$.

For the beams we have
\begin{equation}
\frac{2 q_a\cdot p_k}{Q_a} = \frac{Q}{Q_a} \abs{\vec p_{kT}}\,e^{Y -\eta_k}
\,,\end{equation}
with $Y-\eta_k \to -Y + \eta_k$ for $a \to b$. Here two potential choices for $Q_{a,b}$ are $Q_{a,b} = Q$,
giving the invariant-mass distance measure, or $Q_{a,b} = Q e^{\pm Y} =
x_{a,b} \Ecm$, which gives
\begin{equation}
\frac{2 q_{a,b}\cdot p_k}{Q_{a,b}} = \abs{\vec p_{kT}} e^{\mp\eta_k}
\,.\end{equation}

We will carry out our analysis and one-loop calculations keeping $Q_i$ arbitrary,
enabling various choices to be explored using our results.  From an
experimental point of view certain choices will be more advantageous than
others. For example, the second choice above for $Q_{a,b}$ is useful if the
total rapidity cannot be measured because there are missing-energy particles in the
final state.

For convenience we define the dimensionless reference momenta and
their invariant products
\begin{equation}
\hq_i^\mu = \frac{q_i^\mu}{Q_i}
\,,\qquad
\hs_{ij} = 2\hq_i\cdot\hq_j
\,.\end{equation}
We can then rewrite \eq{TauN_def} as follows,
\begin{align} \label{eq:TauNi_def}
\Tau_N &= \sum_i \Tau_N^i \equiv \sum_i 2\hq_i \cdot P_i
\nn\\
P_i^\mu &= \sum_k p_k^\mu\, \prod_{j\neq i} \theta\bigl(\hq_j \cdot p_k - \hq_i \cdot p_k \bigr)
\,,\end{align}
where $P_i^\mu$ is the total four-momentum in region $i$. The $\Tau_N^i$ are
thus given by the small light-cone component of the $P_i$ measured along their
respective collinear directions $\hq_i$. In the next section we explore the
factorization theorem that is fully differential in the $\Tau_N^i$. The
resulting fully differential soft function will be the focus of our calculations.

\subsection{\boldmath $N$-Jettiness Differential in Jet Regions}
\label{subsec:diffxsect}

The factorization theorem for $\df\sigma/\df\Tau_N$ was given in
Ref.~\cite{Stewart:2010tn}, and is derived in a straightforward manner from
SCET, see Refs.~\cite{Bauer:2002nz,Bauer:2008jx,Stewart:2009yx} (with an
assumption so far implicit in all $N$-jet factorization formulae about the
cancellation of Glauber gluons).  Instead of measuring $\Tau_N$, the
manipulations leading to the factorization theorem are no more difficult when we
consider the fully differential cross section, where we measure each individual
$\Tau_N^i$. The value of $\Tau_N^i$ determines the transverse mass of region $i$
relative to the direction $\vec n_i$ since
\begin{align}
M_{iT}^2
&= P_i^2 + \vec{P}_{i\perp}^2 = (\bar n_i \cdot P_i) (n_i\cdot P_i)
\nn\\
&= 2 q_i \cdot P_i\, [1 + \ord{\lambda^2}]
\nn\\
&= Q_i \Tau_N^i\, [1 + \ord{\lambda^2}]
\,,\end{align}
where $n_i^\mu = (1,\vec n_i)$, $\bar n_i^\mu = (1,-\vec n_i)$. In the last
line we used $\bar n_i\cdot q_i= \bar n_i \cdot P_i + \ord{\lambda^2}$, where
$\lambda^2 \sim \Tau_N^i/Q$ and the power corrections depend on how the magnitude of
$q_i$ is fixed.

If the label vector $\vec{n}_i$ is chosen to be aligned
with the direction of the jet three-momentum $\vec{P}_{i}$ such that
$\vec{n}_i\cdot \vec{P}_i/\abs{\vec{P}_i} \sim 1 + \ord{\lambda^4}$
then $\vec{P}_{i\perp}^2 = 0 + \ord{\lambda^4}$
and the transverse mass is the same as the invariant mass.
\begin{equation}
M_i^2 = P_i^2 = Q_i \Tau_N^i\, [1 + \ord{\lambda^2}]
\,.\end{equation}
Thus the differential $\Tau_N^i$ spectrum corresponds to the spectrum
in the invariant mass for jet $i$,
where $M_i^2\to 0$ for a pencil like jet of massless partons.

The factorized form for the cross section in the limit where all the $\Tau_N^i$
are assumed to be parametrically comparable but small compared to $Q_i \sim Q$
is
\begin{widetext}
\begin{align} \label{eq:sigma_TauN}
\frac{\df\sigma}{\df \Tau_N^a\, \df\Tau_N^b\dotsb\df\Tau_N^N}
&=
\int\!\df x_a \df x_b \int\!\df^4 q\,\df\Phi_L(q) \int\! \df \Phi_N(\{q_J\})\,
M_N(\Phi_N, \Phi_L)\, (2\pi)^4 \delta^4\bigl(q_a + q_b - q_1 - \dotsb - q_N - q\bigr)
\nn\\ &\quad \times
\sum_{\kappa}
\int\!\df t_a\, B_{\kappa_a}(t_a, x_a, \mu)
\int\!\df t_b\, B_{\kappa_b}(t_b, x_b, \mu)
\prod_{J=1}^N \int\!\df s_J\, J_{\kappa_J}(s_J, \mu)
\\\nn &\quad \times
\vC_N^{\kappa\dagger}(\Phi_N, \Phi_L, \mu)\,
\hS_N^{\kappa} \biggl(\Tau_N^a - \frac{t_a}{Q_a}, \Tau_N^b - \frac{t_b}{Q_b}, \Tau_N^1 - \frac{s_1}{Q_1}, \ldots, \Tau_N^N - \frac{s_N}{Q_N} , \{\hq_i\}, \mu\biggr)\,
\vC_N^{\kappa}(\Phi_N, \Phi_L, \mu)
\,.\end{align}
\end{widetext}
Here, $\Phi_N(\{q_J\})$ denotes the $N$-body massless phase space for the $N$
reference jet momenta $\{q_J\}$, while $\Phi_L(q)$ is the ``leptonic'' phase
space for any additional nonhadronic particles in the final state, whose total
momentum is $q$. The measurement function $M_N(\Phi_N, \Phi_L)$ enforces all $N$
jets to be energetic and well enough separated so that $\hat s_{ij}\gg \Tau_N / Q$.
The index $\kappa$ runs over all relevant partonic channels, with $\kappa_a,
\kappa_b, \ldots, \kappa_N$ denoting the individual parton types.

The hard Wilson coefficient $\vC_N^\kappa$ is a vector in the appropriate color space of the external hard partons in each partonic channel. It only depends on the hard phase-space variables $\Phi_N$ and $\Phi_L$. The soft function $\hS_N^\kappa$ is a matrix in the same color space. We can rewrite the color contraction as
\begin{equation}
\vC_N^\dagger \hS_N \vC_N = \tr (\vC_N \vC_N^\dagger  \hS_N) = \tr (\hH_N \hS_N)
\,,\end{equation}
so the hard function $\hH_N = \vC_N \vC_N^\dagger$ is also a color-space matrix.

We want to compute the $N$-jettiness soft function
\begin{equation} \label{eq:SN}
\hS_N^{\kappa} \bigl(k_a, k_b, k_1, \ldots, k_N, \{\hq_i\}, \mu\bigr)
\,.\end{equation}
The $k_i$ are the soft contributions to the $\Tau_N^i$, so from \eq{TauNi_def} we have
\begin{equation} \label{eq:ki_def}
k_i = 2\hq_i \cdot \biggl[\sum_{k\in\mathrm{soft}} p_k\,
\prod_{j\neq i} \theta\bigl(\hq_j \cdot p_k - \hq_i \cdot p_k \bigr) \biggr]
\,,\end{equation}
where the sum now only runs over soft momenta. As indicated by the second to last argument in \eq{SN}, the soft function still depends on all the reference momenta $\{\hq_i\}$, because they enter in the definition of the measured soft momentum components in \eq{ki_def}. The soft function is defined by the vacuum matrix element
\begin{align}
&\hS_N^{\kappa} \bigl(k_a, k_b, k_1, \ldots, k_N, \{\hq_i\} \bigr)
\\\nn & \quad
= \MAe{0}{\widehat{Y}^{\kappa\dagger}(\{\hq_i\}) \prod_i \delta(k_i - 2\hq_i\cdot \hat P_i)\, \widehat{Y}^{\kappa}(\{\hq_i\})}{0}
\,,\end{align}
where the $\hat P_i$ denotes the momentum operator that picks out the total momentum in region $i$.
Here, $\widehat{Y}^\kappa(\{\hq_i\})$ denotes a product of eikonal Wilson lines in the $\hq_i$ directions in the appropriate path-ordering and color representation of the external partons in the partonic channel $\kappa$. The $\widehat{Y}^{\kappa\dagger}$ and $\widehat{Y}^\kappa$ are matrices multiplied in color space. We take their overall normalization to be such that the tree level result for $\hS_N^\kappa$ is $\id\prod_i \delta(k_i)$, where $\id$ is the color identity operator (see \eq{id} below).

In the following, we will often use the short-hand notation
\begin{equation}
\hS_N^\kappa(\{k_i\}, \mu) \equiv \hS_N^\kappa(k_a, k_b, k_1, \ldots, k_N, \{\hq_i\}, \mu)
\,,\end{equation}
and similarly for other functions that depend on all $k_i$, such as the anomalous dimension and counterterm for the soft function.

\subsection{Soft-function RGE}
\label{subsec:RGE}

To derive the structure of the renormalization-group equation (RGE) and anomalous dimension of the soft function $\hS^\kappa_N(k_a, ..., \mu)$, we can use the fact that the factorized cross section in \eq{sigma_TauN} is independent of the renormalization scale $\mu$. For this purpose we can ignore the phase-space integrals and only consider the last two lines in \eq{sigma_TauN}. To simplify the notation, we suppress the index $\kappa$ and the momentum dependence on the label momenta from here on.

As already mentioned, the hard Wilson coefficient $\vC_N$ is a vector in the color space of the external hard partons, so its anomalous dimension $\hga_N$ is a matrix in color space. For $1$-jettiness (or $e^+e^-$ $3$-jettiness), the external partons are $q^\alpha \bar q^\beta g^a$, so the only possible color structure is $T^a_{\alpha\beta}$ and the color space becomes one-dimensional. For $q^\alpha \bar q^\beta g^a g^b$ there are already three different color structures $\vT^{ab}_{\alpha\beta} = \{(T^a T^b)_{\alpha\beta}, (T^b T^a)_{\alpha\beta}, \delta^{ab} \delta_{\alpha\beta}\}$, and so on.

The hard Wilson coefficient $\vC_N$ from matching QCD onto SCET satisfies the RGE
\begin{equation} \label{eq:CN_RGE}
\mu \frac{\df}{\df\mu} \vC_N(\mu) = \hga_C(\mu)\,\vC_N(\mu)
\,.\end{equation}
Its anomalous dimension has the general form~\cite{Chiu:2008vv,Becher:2009qa}
\begin{align} \label{eq:gammaC}
\hga_C^\dagger(\mu)
&= - \Gamma_\cusp[\alpha_s(\mu)] \sum_{i < j} \bT_i\cdot\bT_j
\nn\\ & \quad
 \times \ln\Bigl[(-1)^{\Delta_{ij}}\frac{2q_i\cdot q_j}{\mu^2} - \img 0\Bigr]
  + \hga_C^\dagger[\alpha_s(\mu)]
\,,\end{align}
where we define $\Delta_{ij} = 1$ if $i$ and $j$ are both incoming or both outgoing partons and  $\Delta_{ij} = 0$ if one of them is incoming and the other one outgoing, and with our conventions $q_i\cdot q_j$ is always positive. Here, $\Gamma_\cusp(\alpha_s)$ is the universal cusp anomalous dimension~\cite{Korchemsky:1987wg},
\begin{equation}
\Gamma_\cusp(\alpha_s) = \frac{\alpha_s}{4\pi}\, 4 + \ord{\alpha_s^2}
\,.\end{equation}

The $\bT_i^a$ denotes the color charge of the $i$th external particle when coupling to a gluon with color $a$. It acts on the external color space as
\begin{align}
(\bT^a_i\,\vT)_{\dotsb\alpha_i\dotsb} &= T^a_{\alpha_i\beta_i}\, \vT_{\dotsb\beta_i\dotsb}
\,,\nn\\
(\bT^a_i\,\vT)_{\dotsb\alpha_i\dotsb} &= - T^a_{\beta_i\alpha_i}\,\vT_{\dotsb\beta_i\dotsb}
\,,\nn\\
(\bT^a_i\,\vT)^{\dotsb a_i\dotsb} &= \img f^{a_i a b_i}\, \vT^{\dotsb b_i\dotsb}
\,,\end{align}
where the first line is for the $i$th particle being an outgoing quark or incoming antiquark, the second line for an incoming quark or outgoing antiquark, and the third line for a gluon. The product
\begin{equation}
\bT_i\cdot \bT_j = \sum_a \bT_i^a \bT_j^a
\,,\end{equation}
appearing in the first term in \eq{gammaC}, thus represents a particular color-space matrix $\hT_{ij}$ for each choice of $i$ and $j$. We also define the identity operator $\id$, which acts as
\begin{equation} \label{eq:id}
(\id\, \vT)^{\dotsb a_i\dotsb}_{\dotsb \alpha_j \dotsb} = \vT^{\dotsb a_i\dotsb}_{\dotsb \alpha_j \dotsb}
\,.\end{equation}
In particular
\begin{equation}
\bT_i^2 = \id\, C_i
\quad\text{where}\quad
C_q = C_{\bar q} = C_F
\,,\quad
C_g = C_A
\,.\end{equation}
With only three partons, $q^\alpha \bar q^\beta g^a$, the only possible color structure is $T^a_{\alpha\beta}$, so in this case the color matrices are just numbers,
\begin{align}
\id  = 1
\,,\qquad
\bT^2_q = \bT^2_{\bar{q}} &= C_F
\,,\qquad
\bT^2_g = C_A
\,,\nn \\
\bT_q \cdot \bT_{\bar{q}} &=  \frac{C_A}{2} - C_F
\,,\nn \\
\bT_q \cdot \bT_g = \bT_{\bar{q}} \cdot \bT_g &= - \frac{C_A}{2}
\,.\end{align}

Up to two loops~\cite{MertAybat:2006mz} (and maybe more~\cite{Gardi:2009qi, Becher:2009cu, Becher:2009qa, Dixon:2009ur}), the non-cusp term, $\hga_C(\alpha_s)$, in \eq{gammaC} is diagonal in color and given by a sum over individual quark and gluon contributions,
\begin{align}
\hga_C(\alpha_s) &= \id \sum_i \gamma_C^i(\alpha_s) + \ord{\alpha_s^3}
\,,\nn\\*
\gamma_C^q(\alpha_s) = \gamma_C^{\bar q}(\alpha_s)
&= - \frac{\alpha_s}{4\pi}\, 3 C_F + \ord{\alpha_s^2}
\,,\nn\\*
\gamma_C^g(\alpha_s)
&= - \frac{\alpha_s}{4\pi}\, \beta_0 + \ord{\alpha_s^2}
\,.\end{align}

The RGEs for the beam and jet functions are
\begin{align} \label{eq:BJ_RGE}
\mu\,\frac{\df}{\df\mu} B_i(t,x, \mu)
&= \int\!\df t'\, \gamma^i_B(t - t', \mu)\, B_i(t', x, \mu)
\,,\nn\\
\mu\,\frac{\df}{\df\mu} J_i(s, \mu)
&= \int\!\df s'\, \gamma^i_J(s - s', \mu)\, J_i(s', \mu)
\,.\end{align}
The beam and jet anomalous dimension are identical to all orders~\cite{Stewart:2010qs}, and are given by
\begin{align} \label{eq:gammaBJ}
\gamma_B^i(s, \mu) = \gamma_J^i(s, \mu)
&= -2\, C_i\, \Gamma_\cusp[\alpha_s(\mu)] \,\frac{1}{\mu^2}\cL_0\Bigl(\frac{s}{\mu^2}\Bigr)
\nn\\ &\quad
+ \gamma_J^i[\alpha_s(\mu)]\,\delta(s)
\,,\end{align}
with
\begin{align} \label{eq:gammaBJi}
\gamma_J^q(\alpha_s) = \gamma_J^{\bar q}(\alpha_s)
&= \frac{\alpha_s}{4\pi}\, 6 C_F + \ord{\alpha_s^2}
\,,\nn\\
\gamma_J^g(\alpha_s)
&= \frac{\alpha_s}{4\pi}\, 2\beta_0 + \ord{\alpha_s^2}
\,,\end{align}
and $\cL_n(x)$ denotes the standard plus distribution,
\begin{equation}
\cL_n(x) = \biggl[\frac{\theta(x)\ln^n\!x}{x}\biggr]_+
\,.\end{equation}

Taking the derivative of \eq{sigma_TauN} with respect to $\mu$, we now require
\begin{align} \label{eq:mu_cond}
0 &= \mu\frac{\df}{\df\mu} \!\int \Bigl[\prod_i\df s_i\, \cJ_i(s_i, \mu)\Bigr]
\nn\\ & \quad\times
\vC_N^\dagger(\mu)\, \hS_N \Bigl(\Bigl\{\Tau_N^i - \frac{s_i}{Q_i} \Bigr\}, \mu \Bigr) \vC_N(\mu)
\,,\end{align}
where we use $\cJ_i(s_i, \mu)$ to denote either beam or jet functions (with $s_{a,b}\equiv t_{a,b}$), since their RGEs are identical, and as before $i=a,b,1,\ldots, N$. Using \eqs{CN_RGE}{BJ_RGE} together with \eq{mu_cond}, we get
\begin{widetext}
\begin{align}
0
&= \int \Bigl[\prod_i \df s_i\,\df s_i'\, \cJ_i(Q_i\Tau_N^i - s_i', \mu) \Bigr] \biggl\{
\Bigl[\sum_i \gamma_J^i(s_i' - s_i,\mu) \prod_{j\neq i}\delta(s_j' - s_j) \Bigr]
\hS_N \Bigl(\Bigl\{\frac{s_i}{Q_i} \Bigr\}, \mu \Bigr)
\nn\\ & \quad
+ \Bigl[\prod_i \delta(s_i' - s_i) \Bigr]
\Bigl[\hga_C^\dagger(\mu)\, \hS_N\Bigl(\Bigl\{\frac{s_i}{Q_i} \Bigr\}, \mu \Bigr)
+ \hS_N\Bigl(\Bigl\{\frac{s_i}{Q_i} \Bigr\}, \mu \Bigr)\, \hga_C(\mu)
+ \mu\frac{\df}{\df\mu} \hS_N\Bigl(\Bigl\{\frac{s_i}{Q_i} \Bigr\}, \mu \Bigr) \Bigr] \biggr\}
\,,\end{align}
where we divided out the Wilson coefficients and shifted the integration variables $s_i \to Q_i\Tau_N^i - s_i$ and $s_i' \to Q_i\Tau_N^i - s_i'$. We can now multiply by $\prod_i \cJ_i^{-1}(Q_i k_i - Q_i\Tau_N^i, \mu)$ and integrate over $\Tau_N^i$, which replaces $\cJ_i(Q_i \Tau_N^i - s_i', \mu) \to \delta(Q_i k_i - s_i')/Q_i$. Renaming $k_i' = s_i/Q_i$, we obtain
\begin{align} \label{eq:SN_RGE}
\mu\frac{\df}{\df\mu} \hS_N(\{k_i\}, \mu)
&= \int \Bigl[\prod_i \df k_i' \Bigr] \frac{1}{2} \Bigl[ \hga_S(\{k_i - k_i'\}, \mu)\, \hS_N(\{k_i'\}, \mu ) + \hS_N(\{k_i'\}, \mu )\, \hga_S^\dagger(\{k_i - k_i'\}, \mu) \Bigr]
\,,\end{align}
where the soft anomalous dimension is given by
\begin{align} \label{eq:gammaS}
\hga_S(\{k_i\}, \mu)
&= - \id \sum_i Q_i \gamma_J^i(Q_i k_i,\mu) \prod_{j\neq i}\delta(k_j)
- 2 \hga_C^\dagger(\mu) \prod_i \delta(k_i)
\nn\\
&= 2\Gamma_\cusp[\alpha_s(\mu)]  \biggl\{
\sum_i \bT_i^2\,\frac{Q_i}{\mu^2}\cL_0\Bigl(\frac{Q_i k_i}{\mu^2}\Bigr)\prod_{j\neq i}\delta(k_j)
+ \sum_{i < j} \bT_i\cdot\bT_j \ln\Bigl[ (-1)^{\Delta_{ij}} \frac{2q_i\cdot q_j}{\mu^2} - \img 0\Bigr] \prod_i \delta(k_i) \biggr\}
\nn\\ & \quad
- \Bigl\{\id \sum_i \gamma_J^i[\alpha_s(\mu)] + 2 \hga_C^\dagger[\alpha_s(\mu)] \Bigr\} \prod_i \delta(k_i)
\nn\\
&= - 2\Gamma_\cusp[\alpha_s(\mu)]
\sum_{i \neq j} \bT_i\cdot\bT_j \biggr[
\frac{1}{\sqrt{\hs_{ij}}\, \mu}\cL_0\Bigl(\frac{k_i}{\sqrt{\hs_{ij}}\,\mu}\Bigr)
+ \frac{\img \pi}{2} \Delta_{ij} \, \delta(k_i) \biggr]
\prod_{m\neq i} \delta(k_m)
+ \hga_S[\alpha_s(\mu)] \prod_i \delta(k_i)
\,,\end{align}
\end{widetext}
with the non-cusp part
\begin{align} \label{eq:gammaSi}
\hga_S(\alpha_s)
&= -\id \sum_i \gamma_J^i(\alpha_s) - 2 \hga_C^\dagger(\alpha_s)
\nn\\
&= -\id \sum_i \bigr[\gamma_J^i(\alpha_s) + 2 \gamma_C^i(\alpha_s) \bigr]  + \ord{\alpha_s^3}
\nn\\
&= 0 + \ord{\alpha_s^2}
\,.\end{align}
In the last step above we rescaled the plus distribution $\lambda \cL_0(\lambda x) = \cL_0(x) + \ln\lambda\, \delta(x)$, and applied color identities like
\begin{equation} \label{eq:color}
\sum_i x_i \bT_i^2 = - \sum_{i\neq j} x_i \bT_i\cdot \bT_j = - \sum_{i < j}(x_i + x_j) \bT_i\cdot \bT_j
\,,\end{equation}
which follows from color conservation, $\sum_i \bT_i = 0$.

This derivation shows that factorization implies that the kinematic dependence of $\hga_S(\{k_i\}, \mu)$ on $k_i$ is separable into individual contributions to all orders. This generalizes the same result obtained for the special case of the hemisphere (i.e. $e^+ e^-$ $2$-jettiness) soft function in Ref.~\cite{Fleming:2007xt}, which is reproduced by \eq{gammaS} using $\bT_i\cdot\bT_j = - C_F$ and $2\hq_i\cdot \hq_j = 1$.

Since in \eq{gammaS} $\Gamma_\cusp(\alpha_s)$, $\gamma_J(\alpha_s)$, and $\hga_C(\alpha_s)$ are all known to two loops, so is $\hga_S(\{k_i\},\mu)$. The general evolution formula in \eq{SN_RGE} leaves $\hS_N$ hermitian, which from \eq{sigma_TauN} is the only requirement to obtain a real cross section.

\subsection{Renormalization and One-Loop Divergences}

The result for the soft anomalous dimension allows us to infer the one-loop counterterm for the soft function in $\overline{\mathrm{MS}}$, which we will need in our calculation to renormalize the bare soft function. This will provide us with a nontrivial cross check on our calculation.

The structure of the anomalous dimension implies that the bare and $\overline{\mathrm{MS}}$ renormalized soft functions are related by
\begin{align} \label{eq:SN_ren}
\hS^\bare_N(\{k_i\})
&= \int \Bigl[\prod_i \df k_i'\,\df k_i'' \Bigr]
\hZ_S(\{k_i'\}, \mu)
\\\nn & \quad \times
\hS_N(\{k_i - k_i' - k_i''\}, \mu)\, \hZ_S^\dagger(\{k_i''\}, \mu)
\,.\end{align}
The bare soft function is independent of $\mu$, so differentiating both sides with respect to $\mu$ determines the soft anomalous dimension in terms of the counterterm,
\begin{align} \label{eq:ZS}
\hga_S(\{k_i\}, \mu)
&= - 2\int \Bigl[\prod_i \df k_i' \Bigr]
\nn\\ & \quad\times
(\hZ_S)^{-1}(\{k_i - k_i'\}, \mu)\, \mu\frac{\df}{\df\mu} \hZ_S(\{k_i'\}, \mu)
\nn\\
&= -2\,\mu\frac{\df}{\df\mu} \hZ_S(\{k_i'\}, \mu) + \ord{\alpha_s^2}
\,.\end{align}
Using \eqs{gammaS}{gammaSi}, the NLO counterterm is thus given by
\begin{align} \label{eq:ZSone}
&\hZ_S(\{k_i\}, \mu)
\nn\\ & \quad
= \id \prod_i \delta(k_i)
- \frac{\alpha_s(\mu)}{2\pi}\,\frac{1}{\eps} \sum_{i\neq j} \bT_i\cdot\bT_j \biggl[
-\frac{1}{2\eps}\, \delta(k_i)
\nn\\ & \qquad
+  \frac{1}{\sqrt{\hs_{ij}}\, \mu}\cL_0\biggl(\frac{k_i}{\sqrt{\hs_{ij}}\,\mu}\biggr)
+  \frac{\img \pi}{2} \Delta_{ij}\, \delta(k_i)
\biggr] \prod_{m\neq i} \delta(k_m)
\nn\\ & \qquad
+ \ord{\alpha_s^2}
\,.\end{align}
Note that since $\hS_N$ is color diagonal at tree level, the imaginary part of $\hZ_S$ does not contribute in \eq{SN_ren} at NLO, because it cancels between $\hZ_S$ and $\hZ_S^\dagger$. Hence, from \eq{ZSone} we expect the UV-divergent parts of the one-loop bare soft function, $\hS^{\bare\one}_N$, to have the form
\begin{align} \label{eq:Sbare_expected}
\hS^{\bare\one}_N(\{k_i\})
&= -\frac{\alpha_s(\mu)}{\pi} \frac{1}{\eps} \sum_{i\neq j} \bT_i\cdot \bT_j\,
\frac{(\hs_{ij}\,\mu^2)^\eps}{k_i^{1 + 2\eps}} \!\prod_{m\neq i} \delta(k_m)
\nn\\ & \quad
+ \ord{\eps^0}
\,.\end{align}
This implies that the UV divergences are given for any $N$ by a simple sum over individual hemisphere contributions. We will see how this happens explicitly in the next section.

\section{\boldmath NLO Calculation}
\label{sec:1jet}

\begin{figure}[t!]
\subfigure[]{\includegraphics[scale=0.63]{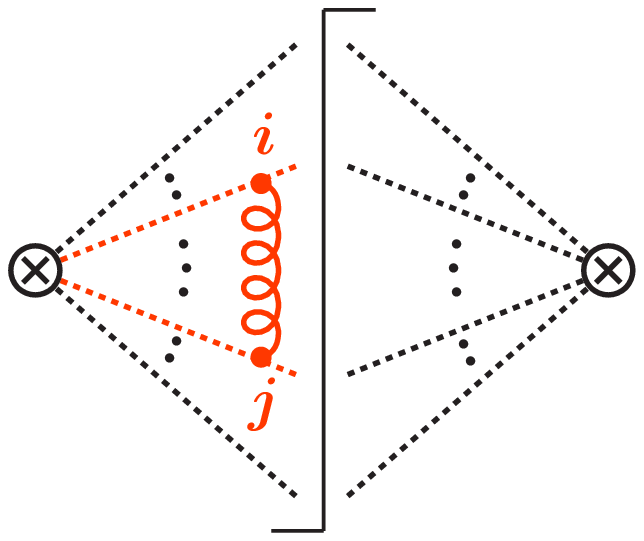}\label{fig:SNvirtual}}%
\hfill%
\subfigure[]{\includegraphics[scale=0.63]{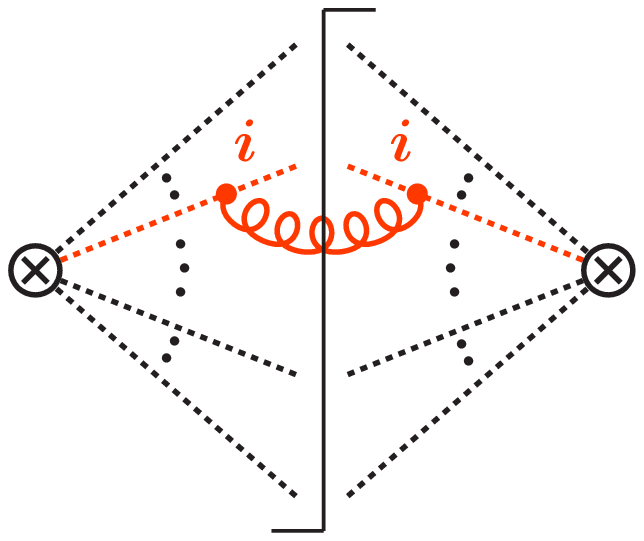}\label{fig:SNii}}%
\\
\subfigure[]{\includegraphics[scale=0.63]{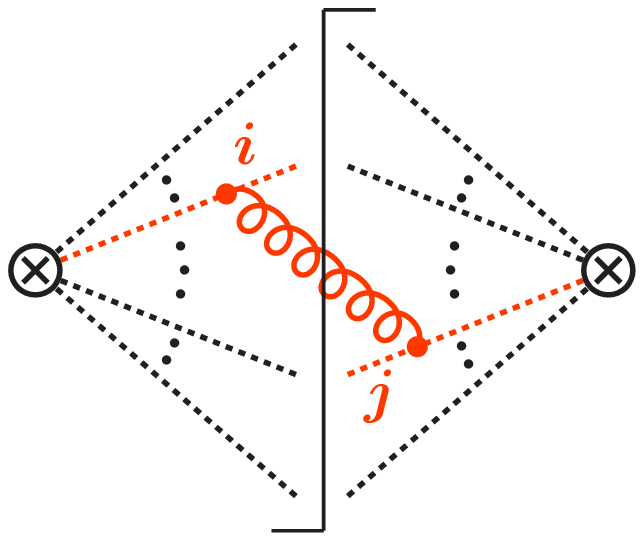}\label{fig:SNij}}%
\hfill%
\subfigure[]{\includegraphics[scale=0.63]{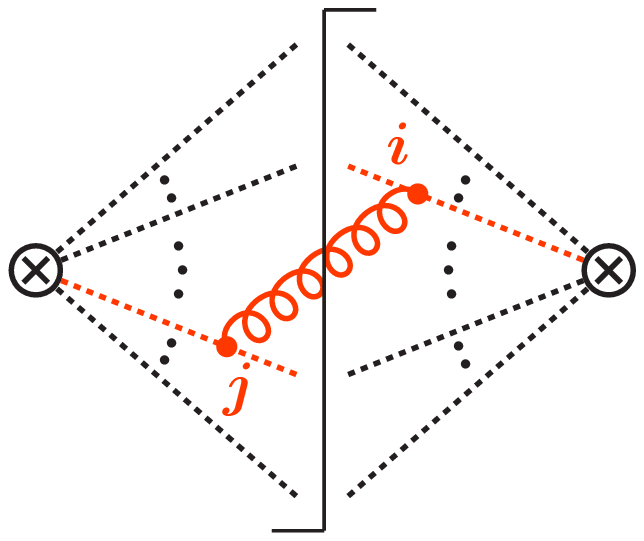}\label{fig:SNji}}%
\vspace{-0.5ex}
\caption{One-loop diagrams for $\hS_N$. The vertical line denotes the final-state cut. Diagrams (a) and (b) vanish. Diagrams (c) and (d) yield \eq{Sijbare_general}.}
\label{fig:SNoneloop}
\end{figure}

\subsection{General Setup}
\label{subsec:setup}

\begin{figure*}[ht!]
\includegraphics[width= \textwidth]{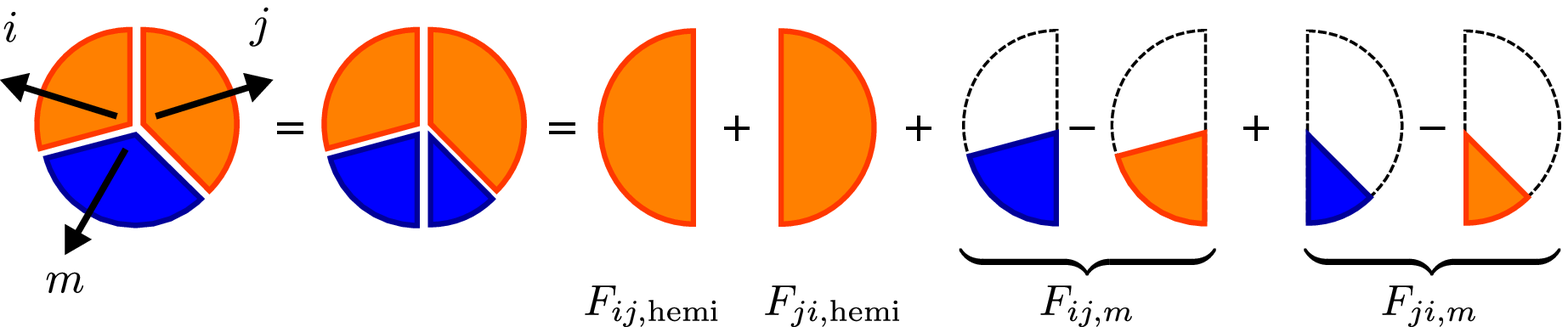}%
\vspace{-0.5ex}
\caption{\label{fig:phasespace} Separation of the measurement function into hemisphere and non-hemisphere measurement functions for $1$-jettiness or $e^+e^-$ $3$-jettiness for a gluon emitted from the $i$th and $j$th Wilson line. The phase space is divided into $i$ and $j$ hemispheres into which the third jet region $m$ is split.}
\end{figure*}

Our calculation in the following applies to both hadronic and $e^+e^-$ collisions, i.e.\ it is independent of whether the Wilson lines are in- or outgoing. For simplicity we use the term ``jet'' to refer to both beam jets and final-state jets.

The one-loop diagrams for the soft function are shown in \fig{SNoneloop}, where $i$ and $j$ label any two Wilson lines, and we work in Feynman gauge. The virtual diagrams in \fig{SNvirtual} are scaleless and thus vanish in pure dimensional regularization. The real emission diagrams in \fig{SNii} with the gluon attaching to the same Wilson line vanish, as it is proportional to $\hq_i^2 = 0$. Hence, at one loop we can write the bare soft function as a sum over contributions where the intermediate gluon attaches to the $i$th and $j$th Wilson line as shown in \figs{SNij}{SNji},
\begin{align} \label{eq:Sijbare_general}
&\hS_N^{\bare\one}(\{k_i\})
\nn\\ & \quad
= -2 \sum_{i < j}\, \bT_i\cdot \bT_j\,
\Bigl(\frac{e^{\gamma_E}\mu^2}{4\pi}\Bigr)^\eps g^2
\int\! \frac{\df^d p}{(2\pi)^d}\,\frac{\hq_i\cdot\hq_j}{(\hq_i\cdot p)(\hq_j\cdot p)}\,
\nn\\ & \qquad \times
2\pi \delta(p^2)\,\theta(p^0)\, F(\{k_i\}, \{2\hq_i \cdot p\})
\,.\end{align}

The key idea in the hemisphere decomposition method is to first fix $i$ and $j$ and then analyze the remaining integral. The measurement function resulting from \eq{ki_def}, which measures the contribution of the gluon in the final state to $k^i$, is given by
\begin{equation} \label{eq:F_def}
F(\{k_i\}, \{p^i\}) = \sum_m \delta(k_m - p^m) \prod_{l\neq m} \delta(k_l)\, \theta(p^l - p^m)
\,,\end{equation}
where we denote the component of the gluon momentum $p^\mu$ along the jet direction $\hq_i^\mu$ as
\begin{equation} \label{eq:pcomp}
p^i = 2\hq_i\cdot p
\,.\end{equation}

For example, for $1$-jettiness (or $e^+e^-$ $3$-jettiness) we have three independent labels $i\neq j\neq m$, so
\begin{align}
&F(\{k_i\}, \{p^i\})
\\\nn & \quad
= \delta(k_i - p^i)\, \delta(k_j)\, \delta(k_m)\, \theta(p^j - p^i)\, \theta(p^m - p^i)
\nn\\ & \qquad
+ \delta(k_i)\, \delta(k_j - p^j)\, \delta(k_m)\, \theta(p^i - p^j)\, \theta(p^m - p^j)
\nn\\\nn & \qquad
+ \delta(k_i)\, \delta(k_j)\, \delta(k_m - p^m)\, \theta(p^i - p^m)\, \theta(p^j - p^m)
\,.\end{align}
The first two terms correspond to the case where the gluon emitted from the $i$th and $j$th Wilson line ends up in the region of jet $i$ or jet $j$, respectively. In this case, $p$ can become collinear with either $\hq_i$ or $\hq_j$, resulting in a double UV-IR divergence. In the last term, the gluon is in the remaining jet $m$. In this case, both $p^i > p^m$ and $p^j > p^m$ are bounded from below, so there is only a single soft IR divergence. The virtual diagrams, which vanish in pure dimensional regularization, turn all IR divergences into UV divergences.

To combine the divergences from the different jet regions, we split the region of jet $m$ into the two hemispheres $p^i < p^j$ and $p^i > p^j$ defined by the directions of jets $i$ and $j$,
\begin{align} \label{eq:F_split}
&F(\{k_i\}, \{p^i\})
\nn\\* & \quad
= \theta(p^j - p^i) \Bigl[\delta(k_i - p^i)\, \delta(k_m)\, \theta(p^m - p^i)
\nn\\* & \qquad
+ \delta(k_i)\, \delta(k_m - p^m)\, \theta(p^i - p^m)
\Bigr] \delta(k_j) + (i\leftrightarrow j)
\nn\\ & \quad
= \delta(k_i - p^i)\, \delta(k_j)\, \theta(p^j - p^i)\, \delta(k_m)
\nn\\ & \qquad
+ \Bigl[\delta(k_i)\, \delta(k_m - p^m)- \delta(k_i - p^i)\, \delta(k_m)\Bigr]\,
\nn\\ & \qquad \quad\times
\delta(k_j)\, \theta(p^j - p^i) \, \theta(p^i - p^m)
+ (i\leftrightarrow j)
\nn\\ & \quad
\equiv F_{ij,\hemi}(\{k_i\}, \{p^i\}) + F_{ji,\hemi}(\{k_i\}, \{p^i\})
\nn\\ & \qquad
+ F_{ij,m}(\{k_i\}, \{p^i\}) + F_{ji,m}(\{k_i\}, \{p^i\})
\,.\end{align}
In the second step we replaced $\theta(p^m - p^i) = 1 - \theta(p^i - p^m)$ in the first term to extend the region for jet $i$ to the full $p^i < p^j$ hemisphere, which gives the hemisphere measurement function
\begin{equation}
F_{ij,\hemi}(\{k_i\}, \{p^i\}) = \delta(k_i - p^i)\, \delta(k_j)\, \theta(p^j - p^i)\, \delta(k_m)
\,.\end{equation}
The contribution for $p^m < p^i < p^j$, which overlaps with the region for jet $m$, is subtracted in the second term, which gives the non-hemisphere measurement function for region $m$,
\begin{align} \label{eq:Fijm}
&F_{ij,m}(\{k_i\}, \{p^i\})
\nn\\ & \qquad
= \bigl[\delta(k_i)\, \delta(k_m - p^m)- \delta(k_i - p^i)\, \delta(k_m)\bigr]\,
\nn\\ & \qquad\quad\times
\delta(k_j)\, \theta(p^j - p^i) \, \theta(p^i - p^m)
\,.\end{align}
This splitting up of the measurement function is illustrated in \fig{phasespace}. The integrations over the $i$ and $j$ hemispheres resulting from $F_{ij,\hemi}$ and $F_{ji,\hemi}$ will now contain all divergences, while the integration over region $m$ resulting from $F_{ij,m}$ and $F_{ji,m}$ will be UV and IR finite, as we will see explicitly below. Essentially, the restriction of the emitted and measured gluon to stay away from the $i$ and $j$ directions, $p^{i,j} > p^m$, cuts off the UV divergence, while the subtraction of the overlapping hemisphere contribution removes the soft divergence: In the soft limit, both $p^i\to 0$ and $p^m\to 0$, and the two terms in square brackets in \eq{Fijm} cancel each other.

We will see in \subsec{Njet} (see \eq{FGN_split} below) that this split up of the measurement function generalizes to any $N$. Hence, we write the renormalized soft function as
\begin{align} \label{eq:Sij_def}
\hS_N(\{k_i\}, \mu)
&= \id \prod_i \delta(k_i) + \sum_{i\neq j} \bT_i\cdot \bT_j\, S_{ij}^\one(\{k_i\}, \mu)
\nn\\ & \quad
+ \ord{\alpha_s^2}
\,,\end{align}
where we split the NLO contribution into a hemisphere and a non-hemisphere contribution, with the latter given by a sum over the different jet regions $m \neq i,j$,
\begin{equation} \label{eq:Sij_split}
S_{ij}^\one(\{k_i\}, \mu) = S_{ij,\hemi}^\one(\{k_i\}, \mu) + \sum_{m \neq i,j} S_{ij,m}^\one(\{k_i\}, \mu)
\,.\end{equation}
Here, $S_{ij,\hemi}$ and $S_{ij,m}$ are the contributions corresponding to $F_{ij,\hemi}$ and $F_{ij,m}$ in \eq{F_split}.

It is instructive to compare our hemisphere decomposition with the method used in Ref.~\cite{Ellis:2010rwa} to calculate the soft function for cone jets. There the authors first specify a jet region and then sum over all contributions from different gluon attachments for that fixed jet region. In the end they sum over the different jet regions. In this case there are nontrivial cancellations between the divergences (and finite terms) arising from the same gluon attachment contributing to different jets. In contrast, as seen from \eqs{Sij_def}{Sij_split}, in the hemisphere decomposition we first specify a gluon attachment $i,j$ and then sum over the contributions to the different jet regions $m$ from this specific attachment. This allows us to make the cancellations explicit and to isolate the UV divergences into the hemisphere contributions. In the end we sum
over all possible attachments.

\subsection{Hemisphere Contributions}
\label{subsec:divergent}

Using \eq{Sijbare_general} and restricting the measurement function to the hemisphere contribution, $F_{ij,\hemi}$ in \eq{F_split}, we obtain
\begin{widetext}
\begin{align} \label{eq:Sijbare_hemi}
S_{ij,\hemi}^{\bare\one}(\{k_i\})
&= -2\Bigl(\frac{e^{\gamma_E}\mu^2}{4\pi}\Bigr)^\eps g^2
\int\! \frac{\df^d p}{(2\pi)^d}\,\frac{2\hs_{ij}}{p^i\, p^j}\,
2\pi \delta(p^2)\,\theta(p^0)\, \delta(k_i - p^i)\,  \delta(k_j)\, \theta(p^j - p^i)\, \delta(k_m)
\nn\\
&= - \frac{\alpha_s(\mu)}{\pi}\,\frac{(e^{\gamma_E})^\eps}{\Gamma(1-\eps)}
\bigl(\hs_{ij}\, \mu^2 \bigr)^\eps
\int\!\df p^i\,\df p^j \,\frac{\theta(p^i)\, \theta(p^j)}{(p^i p^j)^{1+\eps}}\,
\delta(k_i - p^i)\,  \delta(k_j)\, \theta(p^j - p^i)\, \delta(k_m)
\nn\\
&= - \frac{\alpha_s(\mu)}{\pi} \frac{1}{\eps}\,\frac{(e^{\gamma_E})^\eps}{\Gamma(1-\eps)}
\bigl(\hs_{ij}\, \mu^2\bigr)^\eps
\frac{\theta(k_i)}{k_i^{1+2\eps}}\, \delta(k_j)\, \delta(k_m)
\,.\end{align}
In the second step we used the coordinate decomposition
\begin{equation}
p^\mu  =  p^j\, \frac{\hq_i^\mu}{\hs_{ij}} +  p^i \frac{\hq_j^\mu}{\hs_{ij}}  +  p_{ij\perp}^\mu
\end{equation}
to rewrite the phase-space integral as
\begin{equation}
\int\!\frac{\df^d p}{(2\pi)^d}\, 2\pi \delta(p^2)\,\theta(p^0)
= \frac{(4\pi)^\eps}{(2\pi)^2\Gamma(1-\eps)}\, \frac{1}{4\hs_{ij}} \int\!\df p^i\,\df p^j
\Bigl(\frac{\hs_{ij}}{p^i\, p^j}\Bigr)^{\!\eps}\, \theta(p^i)\,\theta(p^j)
\,,\end{equation}
and in the last step we integrated over $p^i$ and $p^j$. The result in \eq{Sijbare_hemi} has the expected form in \eq{Sbare_expected} and reproduces the correct counterterm and anomalous dimension. Expanding \eq{Sijbare_hemi} and subtracting the $1/\eps$ divergences, we obtain the renormalized NLO hemisphere contribution
\begin{equation} \label{eq:Sijhemi}
S_{ij,\hemi}^\one(\{k_i\}, \mu)
= \frac{\alpha_s(\mu)}{4\pi}
\biggl[\frac{8}{\sqrt{\hs_{ij}}\, \mu}\cL_1\biggl(\frac{k_i}{\sqrt{\hs_{ij}}\,\mu}\biggr) - \frac{\pi^2}{6}\,\delta(k_i)\biggr]\delta(k_j)\, \delta(k_m)
\,.\end{equation}
This generalizes the one-loop result for the hemisphere soft function for two back-to-back jets with equal energies from Refs.~\cite{Schwartz:2007ib,Fleming:2007xt} to general hemispheres defined by two jet directions $\hq_i$ and $\hq_j$. Note that, as expected from reparametrization invariance, the dependence on the jet directions only appears through the invariant $\hs_{ij}$.

\subsection{Non-Hemisphere Contributions}
\label{subsec:finite}

We now turn to the non-hemisphere contributions that account for the precise definition of the $1$-jettiness observable and the fact that the boundaries between the different jet regions are more complicated than simple hemispheres. Inserting the second part $F_{ij,m}$ in \eq{F_split} into \eq{Sijbare_general}, we get
\begin{align} \label{eq:Sijmbare}
S_{ij,m}^{\bare\one}(\{k_i\})
&= -\frac{\alpha_s(\mu)}{\pi}\, \bigl(e^{\gamma_E}2\hq_i\cdot\hq_j \mu^2 \bigr)^\eps
\int\! \frac{\df\Omega_{d-2}}{2\pi^{1-\eps}}\, \df p^i\,\df p^j\,\frac{\theta(p^i)\, \theta(p^j)}{(p^i p^j)^{1+\eps}}\,
\nn\\* & \quad\times
\Bigl[\delta(k_i)\, \delta(k_m - p^m) - \delta(k_i - p^i)\, \delta(k_m)\Bigr] \delta(k_j)\, \theta(p^j - p^i) \, \theta(p^i - p^m)
\,.\end{align}
To perform the integration over $p^i$ we use the rescaled variable $x = p^j/p^i$, and rewrite $p^m$ in terms of $p^i$, $x$, and the angle $\phi$ between $\vec{q}_{m\perp}$ and $\vec{p}_\perp$ in the transverse plane,
\begin{equation}
x = \frac{p^j}{p^i}
\,,\qquad
\frac{p^m}{p^i} = \frac{\hs_{jm}}{\hs_{ij}} + \frac{\hs_{im}}{\hs_{ij}}\,x
- 2\Bigl(\frac{\hs_{jm}\hs_{im}}{\hs_{ij}^2} x\Bigr)^{1/2} \cos\phi
\equiv z(x, \phi)
\,.\end{equation}
The limit $p^m < p^i$ thus implies an upper limit on $x$, which eliminates the UV divergence for $x\to\infty$. In addition, since $p^m$ scales like $p^i$, there is also no IR divergence in \eq{Sijmbare}, because the term in square brackets vanishes in the limit $p^i \to 0$. The integral over $p^i$ can then be performed without encountering any divergences,
\begin{equation}
\mu^{2\eps} \int\!\df p^i\, \frac{\theta(p^i)}{(p^i)^{1+2\eps}} \Bigl[\delta(k_i)\, \delta(k_m - p^m) - \delta(k_i - p^i)
\delta(k_m)\Bigr]
= \delta(k_i)\,\frac{1}{\mu}\cL_0\Bigl(\frac{k_m}{\mu}\Bigr) - \frac{1}{\mu}\cL_0\Bigl(\frac{k_i}{\mu}\Bigr)\, \delta(k_m) - \ln[z(x, \phi)]\,\delta(k_i)\, \delta(k_m)
\,.\end{equation}
Note that the $\mu$-dependence cancels between the first two terms. Taking $\eps \to 0$ everywhere else, we obtain the NLO non-hemisphere contribution
\begin{align} \label{eq:Sijm}
S_{ij,m}^\one(\{k_i\}, \mu)
&= \frac{\alpha_s(\mu)}{\pi}
\biggl\{
I_0\Bigl(\frac{\hs_{jm}}{\hs_{ij}}, \frac{\hs_{im}}{\hs_{ij}} \Bigr)
\biggl[\frac{1}{\mu}\cL_0\Bigl(\frac{k_i}{\mu}\Bigr)\, \delta(k_m) - \delta(k_i)\,\frac{1}{\mu}\cL_0\Bigl(\frac{k_m}{\mu}\Bigr) + \ln\frac{\hs_{jm}}{\hs_{ij}} \,\delta(k_i)\, \delta(k_m) \biggr] \delta(k_j)
\nn\\ & \quad
+ I_1\Bigl(\frac{\hs_{jm}}{\hs_{ij}}, \frac{\hs_{im}}{\hs_{ij}} \Bigr)\,
\delta(k_i)\,\delta(k_j)\,\delta(k_m) \biggr\}
\,.\end{align}
The remaining finite phase-space integrals are defined as (rescaling $x = y^2 (\alpha/\beta)$ to simplify the integrands)
\begin{align} \label{eq:I01_def}
I_0(\alpha, \beta)
&=  \frac{1}{\pi}\int_{-\pi}^{\pi}\!\df\phi \int\!\frac{\df y}{y}\,
\theta\bigl(y - \sqrt{\beta/\alpha}\bigr)\,
\theta\bigl(1/\alpha - 1 - y^2 + 2 y\cos\phi\bigr)
\,,\nn\\
I_1(\alpha, \beta)
&= \frac{1}{\pi}\int_{-\pi}^{\pi}\!\df\phi \int\!\frac{\df y}{y}\,
\ln(1 + y^2 - 2y\cos\phi\bigr)\,
\theta\bigl(y - \sqrt{\beta/\alpha}\bigr)\, \theta\bigl(1/\alpha - 1 - y^2 + 2y\, \cos\phi\bigr)
\,.\end{align}
\end{widetext}
In \eq{Sijm} they are evaluated at $\alpha = \hs_{jm}/ \hs_{ij}$ and $\beta = \hs_{im}/ \hs_{ij}$.
Their numerical evaluation for fixed $\alpha > 0$ and $\beta > 0$ poses no problem.
We were not able to find complete analytic expressions. Their analytic simplification to one-dimensional integrals is given in \app{integrals}, with the final result in \eq{I01_final}.

\subsection{Extension to Other Observables}
\label{subsec:otherobs}

As we have just seen, we can extract the divergences in the soft function by dividing up phase space into hemispheres corresponding to pairs of Wilson lines. We will now generalize this decomposition to general IR-safe observables and to more than three regions (in which case the $\hq_i$ are in general non-planar).

Consider a measurement that specifies a way to split up the angular phase space into non-overlapping regions. We use the notation $\Theta_i(p) = 1$ when the momentum $p$ is inside region $i$ and $\Theta_i(p) = 0$ otherwise. We require that the union of all regions covers all of phase space, and that each region contains at most one of the directions $\hq_i$, i.e.\
\begin{equation} \label{eq:regions}
\sum_i \Theta_i (p) = 1
\,,\qquad
\Theta_i(\hq_j) = \delta_{ij}
\,.\end{equation}
We explicitly allow the possibility that there are regions that do not contain any of the $\hq_i$, in which case there will be more than $N+2$ regions.

In general, we can measure a different observable in each region. At NLO, we only need to know how the observable for each region $i$ acts on a one-particle state with momentum $p$, which we denote by $f_i(p)$. We want $f_i$ to be IR safe, which implies that $f_i(p\to 0)$ is equivalent to measuring no gluon at all. Without loss of generality we can assume that $f_i(0) = 0$. We will continue to denote the arguments of the soft function by $k_i$, which are now given by the soft contribution to $f_i$. With this notation, the generalization of the measurement function in \eq{F_def} acting on a soft gluon with momentum $p$ is
\begin{align} \label{eq:FG_def}
F(\{k_i\}, p ) &= \sum_m  \delta[ k_m - f_m (p)]\, \Theta_m (p) \prod_{l\neq m} \delta(k_l)
\,.\end{align}

We now want to generalize \eq{F_split} by splitting up \eq{FG_def} into hemisphere and non-hemisphere contributions according to which Wilson lines the gluon attaches to. We continue to use the labels $i$ and $j$ for the directions of these two Wilson lines. The hemispheres are still determined via the gluon momentum components $p^{i,j} = 2\hq_{i,j}\cdot p$ by $p^j > p^i$ and $p^i > p^j$. Writing out \eq{FG_def} we now have
\begin{align}
&F(\{k_i\}, p)
= \theta(p^j - p^i) \Bigl[\delta[k_i - f_i(p)]\,  \prod_{l\neq i} \delta(k_l) \Theta_i(p)
\nn\\ & \qquad
+ \sum_{m\neq i}  \delta[k_m - f_m(p)]\,\prod_{l\neq m} \delta(k_l) \, \Theta_m(p) \Bigr]
+ (i\leftrightarrow j)
\,.\end{align}
Note that region $j$ is allowed to overlap with hemisphere $i$, and vice versa. Using \eq{regions}, we have
\begin{align} \label{eq:regioncomp}
\Theta_i(p)
&= 1 - \sum_{m\neq i} \Theta_m(p)
\,,\end{align}
which allows us to replace the regions $i$ and $j$ by full hemispheres analogous to \eq{F_split}, where the complement of $\Theta_i(p)$ is now split up between the remaining $\Theta_m(p)$ with $m\neq i$. Then \eq{FG_def} can be written as
\begin{align} \label{eq:FGN_split}
F(\{k_i\}, p)
&= F_{ij,\hemi}(\{k_i\}, p) + F_{ji,\hemi}(\{k_i\},  p)
\\ \nn  & \quad
+ \sum_{m\neq i}  F_{ij,m}(\{k_i\},  p)+\sum_{m\neq j} F_{ji,m}(\{k_i\}, p)
\,,\end{align}
where the hemisphere contributions are given by
\begin{equation} \label{eq:FGijhemi}
F_{ij,\hemi}(\{k_i\}, p)
= \theta(p^j -p^i)\, \delta[k_i - f_i(p)]\, \prod_{l\neq i} \delta(k_l)
\,,\end{equation}
and the non-hemisphere contributions by
\begin{align}  \label{eq:FGijm}
&F_{ij,m}(\{k_i\}, p )
\nn  \\  & \quad
= \theta(p^j -p^i)\, \Theta_m (p)\, \prod_{l\neq i,m} \delta(k_l) \\ & \quad
\quad \times \bigl\{\delta(k_i)\, \delta[k_m -f_m(p)]- \delta[k_i - f_i(p)]\, \delta(k_m)\bigr\}
\nn\,.\end{align}

As in \subsec{finite}, all the divergences are contained in the hemisphere contributions, while the non-hemisphere contributions are UV and IR finite. The measurement of either $f_i$ or $f_m$ in \eq{FGijm} fixes the magnitude of $p$, while the restriction of the emitted gluon to region $m$ forces it to stay away from the $i$ and $j$ directions. Taken together this eliminates the UV divergence. The IR-safety of $f_i$ then ensures that in the limit $p \to 0$ the terms in curly brackets in \eq{FGijm} cancel each other, which eliminates the IR divergence. As a result, for any set of IR-safe observables $f_i$ all UV divergences, and hence the anomalous dimension, are contained in the hemisphere contributions determined by \eq{FGijhemi}. Depending on the observable, these contributions can be more complicated than in \eq{Sijhemi}. Note that this result depends on the fact that an observable $f_i$ is measured in each region $i$. If we have a region $u$ where only an angular restriction is imposed by $\Theta_u(p)$, the corresponding $\delta[k_u - f_u(p)]$ is absent (an ``unmeasured jet'' in the language of Ref.~\cite{Ellis:2009wj}). In this case the hemisphere contributions $S_{uj,\hemi}$ are scaleless and vanish. The non-hemisphere contributions $S_{uj,m}$ and $S_{ij,u}$ are still IR-finite, but now contain a UV divergence in the term coming from region $u$, for which the magnitude of $p$ is not fixed anymore. In this case, the factorization structure is different and the soft anomalous dimension depends on the parameters determining the boundary of region $u$, for example the cone radius as in Ref.~\cite{Ellis:2009wj}.

Although we have only applied the hemisphere decomposition method at NLO, the $N$-jettiness factorization theorem implies that the UV divergences and soft anomalous dimensions factor into pairwise hemisphere contributions to all orders, as shown by \eq{gammaS}. Hence, we believe the hemisphere decomposition will remain useful also at higher orders.

\subsection{\boldmath NLO Calculation for $N$-Jettiness}
\label{subsec:Njet}

We now use the general arguments in the previous subsection and apply them to the case of $N$-jettiness. In this case the observables are simply the components of the gluon momentum along the jet directions, while the regions are determined by the smallest $p^i$. Hence,
\begin{equation}
f_i(p) = p^i = 2 \hq_i \cdot p
\,,\qquad
\Theta_i(p) = \prod_{m\neq i} \theta(p^m - p^i)
\,,\end{equation}
which turns \eq{FG_def} into \eq{F_def}. From \eqs{FGijhemi}{FGijm} we get
\begin{equation} \label{eq:FNijhemi}
F_{ij,\hemi}(\{k_i\}, \{p^i\})
= \theta(p^j - p^i)\, \delta(k_i - p^i) \, \prod_{m\neq i} \delta(k_m)
\,,\end{equation}
and
\begin{align} \label{eq:FNijm}
&F_{ij,m}(\{k_i\}, \{p^i\})
\\\nn & \quad
= \bigl[\delta(k_i)\, \delta(k_m - p^m) - \delta(k_i - p^i)\,\delta(k_m) \bigr]
\\\nn &\qquad\times
\theta(p^j - p^i)\, \theta(p^i - p^m) \prod_{l\neq i,m} \delta(k_l)\, \theta(p^l - p^m)
\,.\end{align}

The calculation of the hemisphere contribution for general $N$ is identical to the $1$-jettiness case in \subsec{divergent} with the overall replacement $\delta(k_j)\, \delta(k_m) \to \prod_{m\neq i} \delta(k_m)$ arising from \eq{FNijhemi}. In particular, we can see immediately that this reproduces the correct NLO counterterm and soft anomalous dimension in \eqs{ZSone}{gammaS}. The final result for the renormalized hemisphere contribution is given by \eq{Sijhemi},
\begin{align}
S_{ij,\hemi}^\one(\{k_i\}, \mu)
&= \frac{\alpha_s(\mu)}{4\pi}
\biggl[\frac{8}{\sqrt{\hs_{ij}}\, \mu}\cL_1\biggl(\frac{k_i}{\sqrt{\hs_{ij}}\,\mu}\biggr)
\nn\\ & \quad
 - \frac{\pi^2}{6}\,\delta(k_i)\biggr]
 \prod_{m\neq i}\delta(k_m)
\,.\end{align}

For the non-hemisphere contribution, there are now several regions $m$ contributing. The calculation for each region proceeds as in \subsec{finite}, except that we now have additional $\theta(p^l - p^m)$ functions in \eq{FNijm}, which separate region $m$ from the remaining regions $l \neq i,m$. We can write $p^m$ and $p^l$ in terms of $p^i$ and $x = p^j/p^i$,
\begin{align}
\frac{p^m}{p^i}
&= \frac{\hs_{jm}}{\hs_{ij}} + \frac{\hs_{im}}{\hs_{ij}}\,x
- 2\Bigl(\frac{\hs_{jm}\hs_{im}}{\hs_{ij}^2} x\Bigr)^{1/2} \cos\phi
\,,\nn\\
\frac{p^l}{p^i}
&= \frac{\hs_{jl}}{\hs_{ij}} + \frac{\hs_{il}}{\hs_{ij}}\,x
- 2\Bigl(\frac{\hs_{jl}\hs_{il}}{\hs_{ij}^2} x\Bigr)^{1/2} \cos(\phi + \phi_{lm})
\,.\end{align}
Here $\phi$ is again defined as the angle between $\vec{p}_\perp$ and $\vec{\hq}_{m\perp}$, while $\phi_{lm}$ are the angles between the remaining $\vec{\hq}_{l\perp}$ and $\vec{\hq}_{m\perp}$. The result for $S_{ij,m}^\one(\{k_i\}, \mu)$ has the same form as \eq{Sijm},
\begin{widetext}
\begin{align} \label{eq:SNijm}
S_{ij,m}^\one(\{k_i\}, \mu)
&= \frac{\alpha_s(\mu)}{\pi}
\biggl\{
I_0\Bigl(\frac{\hs_{jm}}{\hs_{ij}}, \frac{\hs_{im}}{\hs_{ij}}, \Bigl\{\frac{\hs_{jl}}{\hs_{jm}}, \frac{\hs_{il}}{\hs_{im}}, \phi_{lm} \Bigr\}_{l\neq i,j,m} \Bigr)
\biggl[
\frac{1}{\mu}\cL_0\Bigl(\frac{k_i}{\mu}\Bigr)\, \delta(k_m) - \delta(k_i)\,\frac{1}{\mu}\cL_0\Bigl(\frac{k_m}{\mu}\Bigr)
\nn\\ & \quad
+ \ln\frac{\hs_{jm}}{\hs_{ij}} \,\delta(k_i)\, \delta(k_m) \biggr]
+ I_1\Bigl(\frac{\hs_{jm}}{\hs_{ij}}, \frac{\hs_{im}}{\hs_{ij}}, \Bigl\{\frac{\hs_{jl}}{\hs_{jm}}, \frac{\hs_{il}}{\hs_{im}}, \phi_{lm} \Bigr\}_{l\neq i,j,m} \Bigr)\,
\delta(k_i)\,\delta(k_m) \biggr\} \prod_{l\neq i,m} \delta(k_l)
\,.\end{align}
The finite phase-space integrals are now given by
\begin{align} \label{eq:I01gen_def}
I_0(\alpha, \beta, \{\alpha_l, \beta_l, \phi_l\})
&=  \frac{1}{\pi}\int_{-\pi}^{\pi}\!\df\phi \int\!\frac{\df y}{y}\,
\theta\bigl(y - \sqrt{\beta/\alpha}\bigr)\, \theta\bigl(1/\alpha - 1 - y^2 + 2 y\cos\phi\bigr)
\nn\\ & \quad\times
\prod_l \theta\Bigl[\alpha_l - 1 + (\beta_l - 1) y^2
- 2y \bigl[\sqrt{\alpha_l \beta_l}\, \cos(\phi +\phi_l) - \cos\phi \bigr] \Bigr]
\,,\nn\\
I_1(\alpha, \beta, \{\alpha_l, \beta_l, \phi_l\})
&= \frac{1}{\pi}\int_{-\pi}^{\pi}\!\df\phi \int\!\frac{\df y}{y}\,
\ln(1 + y^2 - 2y\cos\phi\bigr)\,
\theta\bigl(y - \sqrt{\beta/\alpha}\bigr)\, \theta\bigl(1/\alpha - 1 - y^2 + 2y\, \cos\phi\bigr)
\nn\\ & \quad\times
\prod_l \theta\Bigl[\alpha_l - 1 + (\beta_l - 1) y^2
- 2y \bigl[\sqrt{\alpha_l \beta_l}\, \cos(\phi +\phi_l) - \cos\phi \bigr] \Bigr]
\,.\end{align}
An algorithm to systematically evaluate them numerically is given in \app{integrals}. The values for the parameters in \eq{SNijm} are $\alpha = \hs_{jm}/ \hs_{ij}$, $\beta = \hs_{im}/ \hs_{ij}$, $\alpha_l = \hs_{jl}/ \hs_{jm}$, and $\beta_l= \hs_{il}/ \hs_{im}$.

\end{widetext}

\section{Conclusions}
\label{sec:conclusions}

$N$-jettiness is a global event shape that can be used to define an exclusive $N$-jet cross section. We have given a factorization theorem for the cross section fully differential in the individual $N$-jettiness contributions for each region, $\Tau_N^i$, which correspond to the mass of each jet region. We have computed the corresponding $N$-jettiness soft function, differential in all $\Tau_N^i$ at one loop.

In our calculation we analytically extract the UV divergences by splitting the phase space into hemispheres depending on which Wilson lines the soft gluon attaches to. The hemisphere contributions reproduce the anomalous dimension of the soft function as expected from the consistency of the factorization theorem. The remaining non-hemisphere contributions, which encode the dependence on the boundaries between the regions, are reduced to one-dimensional numerical integrals. We show that this hemisphere decomposition can be applied in general to compute soft functions for other observables, such as jet algorithms and jet shapes, at one loop. We also expect that it can be generalized to two loops.

Our soft-function calculation provides the last missing ingredient to obtain the exclusive $N$-jet cross section resummed to NNLL for any process where the corresponding SCET hard function at NLO is known from the one-loop QCD calculation. In many processes it has been obtained explicitly~\cite{Manohar:2003vb, Bauer:2003di, Idilbi:2005er, Chiu:2008vv, Becher:2009th, Chiu:2009mg, Kelley:2010fn, Ahrens:2010zv, Beneke:2010da}. In general, the NLO hard function is given in terms of the virtual one-loop QCD diagrams, and there are large ongoing efforts to compute these for many LHC processes~\cite{Binoth:2010ra, Campbell:2000bg, Ossola:2007ax, Berger:2008sj, Ellis:2008qc, Bredenstein:2009aj, Bevilacqua:2009zn, Mastrolia:2010nb, Cullen:2011kv}.

The shape of the jet regions as determined by $N$-jettiness depend on the specific distance measure used, and our results apply to any choice of distance measure. As we saw in \fig{2jet_etaphi}, using a geometric measure, the jet regions yield jets with circular boundaries, which is a feature desired experimentally. Hence, it will be interesting to explore the use of $N$-jettiness directly as an exclusive $N$-jet algorithm in the future.

\begin{figure*}
\hspace*{\fill}%
\includegraphics[scale=0.5]{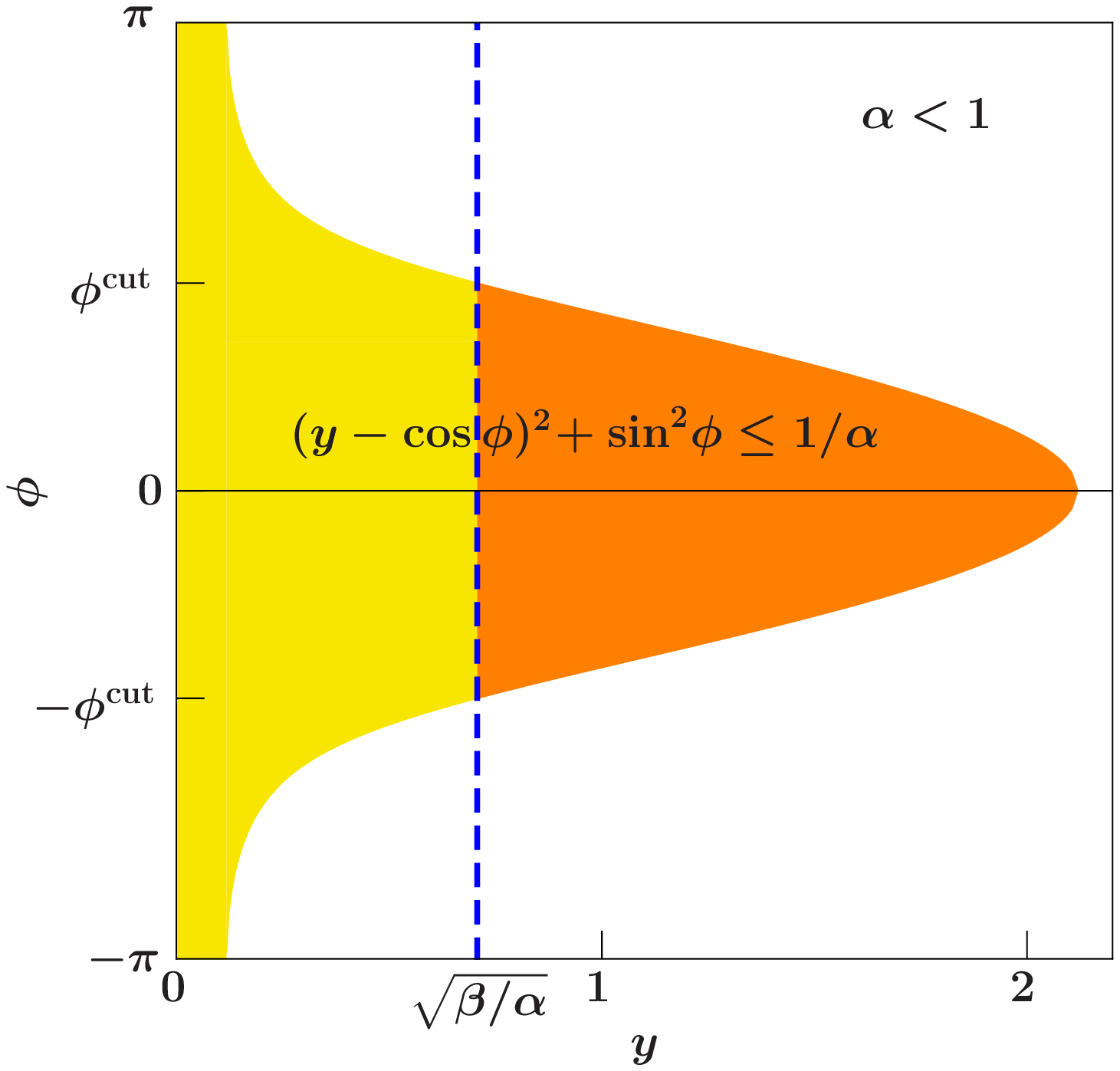}%
\hfill%
\includegraphics[scale=0.5]{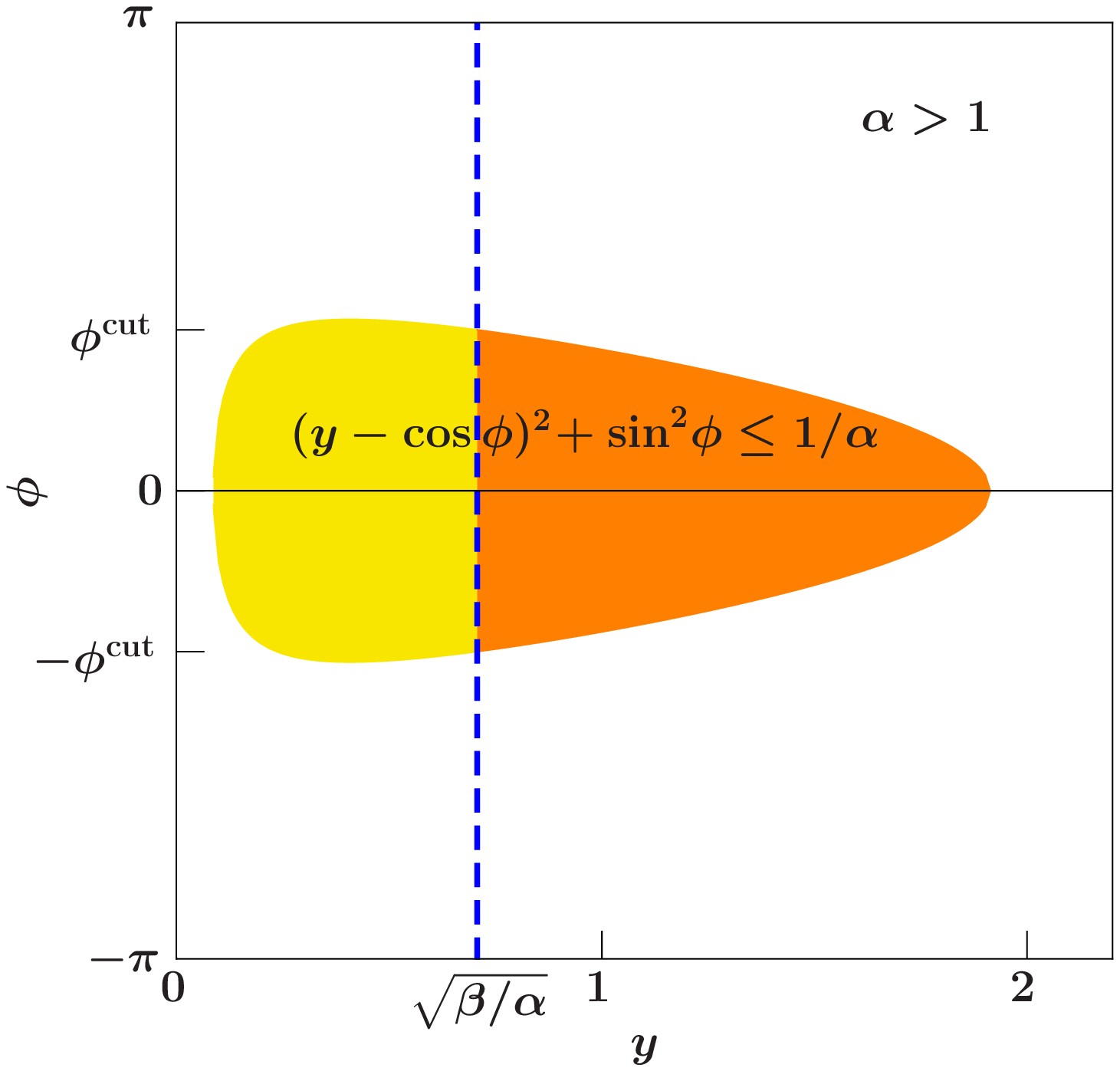}%
\hspace*{\fill}%
\vspace{-0.5ex}
\caption{\label{fig:phiy} Phase-space constraints from \eq{ycondition} in the $\phi$-$y$ plane for $\alpha = 0.8$ (left) and $\alpha = 1.2$ (right). In both cases $\beta/\alpha = 0.5$.}
\end{figure*}

\begin{acknowledgments}
\vspace{-1ex}
This work was supported in part by the Office of Nuclear Physics of the U.S. Department of Energy under the grant DE-FG02-94ER40818, and by the Department of Energy under the grant DE-SC003916. T.J. is also supported by a LHC-TI grant under the NSF grant PHY-0705682.
\end{acknowledgments}

\appendix

\section{Finite Integrals}
\label{app:integrals}

\subsection{\boldmath $1$-Jettiness}

Here we further study the finite phase-space integrals in \eq{I01_def} that are required for $1$-jettiness or $e^+e^-$ $3$-jettiness. The indefinite integrals over $y$ can be carried out explicitly. In particular, for $I_1$ we have
\begin{equation}
G(y, \phi)
= \int\!\frac{\df y}{y} \ln(1 + y^2 - 2y \cos\phi)
= -2 \Re\bigl[\Li_2(y e^{\img\phi})\bigr]
\,.\end{equation}
The remaining integrals over $\phi$ must be be carried out numerically.\footnote{One could also think about first integrating over $\phi$, since the original $\phi$-integral can be done and the limits are linear in $\cos\phi$. This does not lead to any simplification, however, because the remaining numerical $y$-integral will then involve $\arccos[(1+y^2-1/\alpha)/(2y)]$.}

What remains is to determine the $\phi$-dependent integration limits on $y$. We use $-\pi \leq \phi \leq \pi$ as the fundamental region for $\phi$. Also recall that $\alpha = \hs_{jm}/\hs_{ij}$ and $\beta = \hs_{im}/\hs_{ij}$, which are positive definite. The $\theta$ functions in \eq{I01_def} impose the conditions
\begin{equation}
(y - \cos\phi)^2 + \sin^2\!\phi \leq \frac{1}{\alpha}
\,,\qquad
y \geq \sqrt{\frac{\beta}{\alpha}} \geq 0
\,,\end{equation}
which are illustrated in \fig{phiy}. Solving for $y$ they imply
\begin{align} \label{eq:ycondition}
\max\biggl\{\sqrt{\frac{\beta}{\alpha}},\, y_-(\phi, \alpha) \biggr\} &\leq y \leq y_+(\phi, \alpha)
\,,\quad
\sin^2\!\phi \leq \frac{1}{\alpha}
\,,\nn\\
y_-(\phi, \alpha) &= \cos\phi - \sqrt{1/\alpha - \sin^2\!\phi}
\,,\nn\\
y_+(\phi, \alpha) &= \cos\phi + \sqrt{1/\alpha - \sin^2\!\phi}
\,.\end{align}
We can now distinguish the two cases $\alpha \leq 1$ and $\alpha > 1$.

\paragraph*{Case $\alpha \leq 1$}

For $\alpha \leq 1$, we have $\hs_{jm} \leq \hs_{ij}$, which means that $\hq_j$ is closer to $\hq_m$ than to $\hq_i$. In this case, which is illustrated in the left panel of \fig{phiy}, the roots always exist and $y_-(\phi)$ is strictly negative, so we have
\begin{align}
\sqrt{\frac{\beta}{\alpha}} &\leq y \leq y_+(\phi, \alpha)
\,,\nn\\
1 &\geq \cos\phi \geq \max\Bigl\{\frac{\alpha + \beta - 1}{2\sqrt{\alpha\beta}},\, -1\Bigr\}
\,,\nn\\
\sqrt{\beta} &\leq \sqrt{\alpha} + 1
\,.\end{align}
The lower limit on $\cos\phi$ is necessary to guarantee that $\sqrt{\beta/\alpha} \leq y_+(\phi,\alpha)$. The condition on $\alpha$ and $\beta$ is then necessary to guarantee that $1 \geq (\alpha + \beta - 1)/(2\sqrt{\alpha\beta})$, such that the lower $\cos\phi$ limit does not exceed the upper one, otherwise the integral vanishes. The lower $\cos\phi$ limit itself is only nontrivial for $\sqrt{\alpha} + \sqrt{\beta} \geq 1$ which means $\sqrt{\hs_{im}} + \sqrt{\hs_{jm}} \geq \sqrt{\hs_{ij}}$. For a purely geometric measure this is always true, but it need not be the case for more general measures.

\paragraph*{Case $\alpha > 1$}

For $\alpha > 1$, illustrated in the right panel of \fig{phiy}, the condition $\sin^2\phi \leq 1/\alpha$ for the roots to exist becomes nontrivial and forces an upper limit on $\abs{\phi}$,
\begin{equation} \label{eq:philimit}
\abs{\phi} \leq \arcsin\frac{1}{\sqrt{\alpha}}
\,.\end{equation}
(The second solution for the $\arcsin$ is not allowed for $y\geq 0$.) Now, both lower limits on $y$ are possible. To determine which $y$-limit applies at a given value of $\phi$, we can distinguish two cases. First,
\begin{align}
\sqrt{\frac{\beta}{\alpha}} &\leq y \leq y_+(\phi, \alpha)
\,,\nn\\
1 &\geq \cos\phi \geq \frac{\alpha + \beta - 1}{2\sqrt{\alpha\beta}}
\,,\nn\\
\sqrt{\alpha} - 1 &\leq \sqrt{\beta} \leq \sqrt{\alpha} + 1
\,,\end{align}
where the $\cos\phi$ limits result from enforcing $y_- \leq \sqrt{\beta/\alpha} \leq y_+$ and the conditions on $\alpha$ and $\beta$ enforce the lower limit on $\cos\phi$ to be $\leq 1$. Second,
\begin{align}
y_-(\phi, \alpha) &\leq y \leq y_+(\phi, \alpha)
\,,\nn\\
\min \Bigl\{1,\,\frac{\alpha + \beta - 1}{2\sqrt{\alpha\beta}}\Bigr\}
&\geq \cos\phi \geq \sqrt{1 - \frac{1}{\alpha}}
\,,\nn\\
\beta &\leq \alpha - 1
\,,\end{align}
where the upper $\cos\phi$ limit and the condition on $\alpha$ and $\beta$ arises from requiring $\sqrt{\beta/\alpha} \leq y_-(\phi, \alpha)$, while the lower limit on $\cos\phi$ is equivalent to \eq{philimit}.

\paragraph*{Combined Result}

To write the various conditions in a compact form we define the following two angles
\begin{align}
\phi_\mathrm{max}(\alpha) &= \arcsin\frac{1}{\sqrt{\alpha}}
\,,\nn\\
\phi_\cut(\alpha, \beta)
&= \begin{cases}
  0 &\abs{\sqrt{\alpha}-\sqrt{\beta}} \geq 1
  \,,\\
  \pi & \sqrt{\alpha}+\sqrt{\beta} \leq 1
  \,,\\ \displaystyle
  \arccos\frac{\alpha+\beta - 1}{2\sqrt{\alpha\beta}}
  & \text{otherwise}
\,.\end{cases}
\end{align}
The conditions for $\alpha < 1$ and the first case for $\alpha > 1$ reduce to $\abs{\phi} \leq \phi_\cut$. For the second case for $\alpha > 1$, which only applies for $\beta \leq \alpha - 1$, we have $\phi_\cut \leq \abs{\phi} \leq \phi_\mathrm{max}$. Using the fact that the integrand is symmetric in $\phi$, the final result for the integrals is given by
\begin{widetext}
\begin{align} \label{eq:I01_final}
I_0(\alpha, \beta)
&= 2\int_{0}^{\phi_\cut(\alpha, \beta)}\!\frac{\df\phi}{\pi}
\ln\!\frac{y_+(\phi, \alpha)}{\sqrt{\beta/\alpha}}
+ 2\theta(\alpha - \beta - 1)\,
\int^{\phi_\mathrm{max}(\alpha)}_{\phi_\cut(\alpha, \beta)}\!\frac{\df\phi}{\pi}\,
\ln\!\frac{y_+(\phi, \alpha)}{y_-(\phi, \alpha)}
\,,\nn\\
I_1(\alpha, \beta)
&= 2\int_{0}^{\phi_\cut(\alpha, \beta)}\!\frac{\df\phi}{\pi}\,
\bigl[G\bigl(y_+(\phi, \alpha), \phi\bigr) - G\bigl(\sqrt{\beta/\alpha}, \phi\bigr) \bigr]
\nn\\ & \quad
+ 2\theta(\alpha - \beta - 1)\,
\int^{\phi_\mathrm{max}(\alpha)}_{\phi_\cut(\alpha, \beta)}\!\frac{\df\phi}{\pi}\,
\bigl[G\bigl(y_+(\phi, \alpha), \phi\bigr) - G\bigl(y_-(\phi, \alpha), \phi\bigr) \bigr]
\,.\end{align}

\subsection{\boldmath $N$-Jettiness}

We now turn to the integrals $I_{0,1}(\alpha, \beta, \{\alpha_l, \beta_l, \phi_l\})$, defined in \eq{I01gen_def}, that are needed for general $N$. The $y$-integral is the same as before and can be carried out explicitly. For a given value of $\phi$, the $\theta$ functions split the $y$ integration region into a number of mutually exclusive $y$-intervals, which yields
\begin{align} \label{eq:I01gen_final}
I_0(\alpha, \beta, \{\alpha_l, \beta_l, \phi_l\})
&= \int_{-\pi}^{\pi}\!\frac{\df\phi}{\pi}
\sum_{I} \ln\frac{y^I_\mathrm{max}(\phi)}{y^I_\mathrm{min}(\phi)}\,\theta[y_\mathrm{max}^I(\phi) - y_\mathrm{min}^I(\phi)]
\,,\nn\\
I_1(\alpha, \beta, \{\alpha_l, \beta_l, \phi_l\})
&= \int_{-\pi}^{\pi}\!\frac{\df\phi}{\pi}
\sum_{I} \bigl[ G(y^I_\mathrm{max}(\phi), \phi) - G(y^I_\mathrm{min}(\phi), \phi) \bigr]
\,\theta[y_\mathrm{max}^I(\phi) - y_\mathrm{min}^I(\phi)]
\,.\end{align}
Here, the sum runs over all intervals and $y^I_{\mathrm{min}}(\phi)$ and $y^I_{\mathrm{max}}(\phi)$ are the lower and upper limits of the $I$th interval, and can depend on all $\alpha, \beta, \alpha_l, \beta_l, \phi_l$.

What remains is to determine the $y$-limits for a given $\phi$. The conditions imposed by the primary $\theta$ functions involving $\alpha$ and $\beta$ are as in the previous subsection. The additional $\theta$ functions impose the condition for each $l$
\begin{align} \label{eq:lmcondition}
1 - \alpha_l + (1 - \beta_l)\, y^2 - 2 y\bigl[\cos\phi
- \sqrt{\alpha_l  \beta_l}\, \cos(\phi + \phi_l) \bigr]
\leq 0
\,.\end{align}
Recall that $\alpha_l = \hs_{jl}/\hs_{jm} \geq 0$ and $\beta_l = \hs_{il}/\hs_{im} \geq 0$. They essentially compare the distance between $\hq_l$ and $\hq_{i,j}$ with the distance between $\hq_m$ and $\hq_{i,j}$. The angle $\phi_l = \phi_{lm}$ is the angle between $\vec\hq_{l\perp}$ and $\vec\hq_{m\perp}$. The limits on $y$ coming from \eq{lmcondition} are given in terms of the roots of the polynomial,
\begin{equation}
y_{\pm}(\phi, \alpha_l, \beta_l, \phi_l)
= \frac{1}{1 - \beta_l} \biggl\{\cos\phi - \sqrt{\alpha_l  \beta_l} \cos(\phi+ \phi_l)
\pm \sqrt{\bigl[\cos \phi - \sqrt{\alpha_l  \beta_l} \cos(\phi+ \phi_l)\bigr]^2
- (1 - \alpha_l)(1 - \beta_l) } \biggr\}
\,.\end{equation}
\end{widetext}

To analyze the limits on $y$ imposed by \eq{lmcondition} for each $l$, there are three questions to ask:
\begin{enumerate}
\item Does the parabola open upwards or downwards?
\item Does it have real roots?
\item What are the signs of the roots?
\end{enumerate}
The condition for the roots to exist is
\begin{equation} \label{eq:rootcondition}
\bigl[\cos \phi - \sqrt{\alpha_l  \beta_l} \cos(\phi+ \phi_l) \bigr]^2 \geq (1 - \alpha_l)(1 - \beta_l)
\,.\end{equation}
The correct $y$ limits at a given fixed value of $\phi$ are then determined as follows:
\begin{enumerate}

\item $\beta_l < 1$: The parabola opens upwards, so $y$ must be in between the two roots, $y_- \leq y \leq y_+$.

\begin{enumerate}
\item $\alpha_l \geq 1$:
Equation~\eqref{eq:rootcondition} is always satisfied, $y_- \leq 0$, and $y_+ \geq 0$ gives an upper limit
\begin{equation}
y \leq y_+(\phi, \alpha_l, \beta_l, \phi_l)
\,.\end{equation}
\item $\alpha_l < 1$: Equation~\eqref{eq:rootcondition} is nontrivial, and the roots have the same sign if they exist. Hence,
\begin{align}
y_-(\phi, \alpha_l, \beta_l, \phi_l) \leq y &\leq y_+(\phi, \alpha_l, \beta_l, \phi_l)
\,,\nn\\
\cos \phi - \sqrt{\alpha_l  \beta_l} \cos(\phi+ \phi_l) &\geq \sqrt{(1 - \alpha_l)(1 - \beta_l)}
\,.\end{align}
The $y$-integral vanishes if the condition on $\phi$ is not satisfied.
\end{enumerate}

\item $\beta_l > 1$: The parabola opens downwards, so $y$ must be outside the two roots, $y \leq y_-$ or $y \geq y_+$.

\begin{enumerate}
\item $\alpha_l \leq 1$:
Equation~\eqref{eq:rootcondition} is always satisfied, $y_- \leq 0$, and $y_+ \geq 0$ gives lower limit
\begin{equation}
y \geq y_+(\phi, \alpha_l, \beta_l, \phi_l)
\,.\end{equation}
\item $\alpha_l > 1$: Equation~\eqref{eq:rootcondition} is nontrivial, and the roots have the same sign if they exist. Hence,
\begin{align}
y \leq y_-(\phi, \alpha_l, \beta_l, \phi_l)
\quad&\text{or}\quad  y \geq y_+(\phi, \alpha_l, \beta_l, \phi_l)
\,,\nn\\
\cos \phi - \sqrt{\alpha_l  \beta_l} \cos(\phi+ \phi_l) &\geq \sqrt{(1 - \alpha_l)(1 - \beta_l)}
\,.\end{align}
There are no constraints on $y$ if the condition on $\phi$ is not satisfied.
\end{enumerate}

\item $\beta_l = 1$: There is no parabola.

\begin{enumerate}
\item $\alpha_l \leq 1$: The limits are
\begin{align}
y &\geq \frac{1 - \alpha_l}{2\cos\phi - 2\sqrt{\alpha_l }\, \cos(\phi + \phi_l)}
\,,\nn\\
\cos\phi &\geq \sqrt{\alpha_l }\, \cos(\phi + \phi_l)
\,,\end{align}
and the $y$-integral vanishes if the condition on $\phi$ is not satisfied.

\item $\alpha_l > 1$: The limits are
\begin{align}
y &\leq \frac{\alpha_l - 1}{2\sqrt{\alpha_l  }\, \cos(\phi + \phi_l) - 2\cos\phi}
\,,\nn\\
\cos\phi &\leq \sqrt{\alpha_l }\, \cos(\phi + \phi_l)
\,.\end{align}
There are no constraints on $y$ if the condition on $\phi$ is not satisfied.
\end{enumerate}

\end{enumerate}

In principle one can now combine all limits and determine all possible $\phi$-intervals in which a particular set of lower and upper $y$-limits applies, as we did in \eq{I01_final}. However, although this is straightforward it quickly becomes very cumbersome. Alternatively, it is easy to devise an algorithm to obtain the correct $y$-limits in \eq{I01gen_final} for a given value of $\phi$ in the numerical integration over $\phi$. One starts with the $y$-limits in \eq{ycondition}, call them $y_\mathrm{min}$ and $y_\mathrm{max}$. Next, one loops over all $l$ and determines the limits imposed by each $l$ as above. If one encounters a stronger lower or upper limit, $y_\mathrm{min}$ and/or $y_\mathrm{max}$ are updated to the new stronger limit. If one encounters a necessary condition on $\phi$ that is violated, the integrand vanishes and one can stop. Case 2(b) requires special attention. If it is encountered, the $y$ interval is split in two if necessary and one continues by maintaining two (or more) mutually exclusive $y$-intervals each having its own lower and upper limit. Newly encountered stronger limits are then applied to each interval. An interval is eliminated whenever its lower limit exceeds its upper limit. If the last existing interval is eliminated the integrand vanishes.

\bibliographystyle{../physrev4}
\bibliography{../pp}

\end{document}